\documentclass[journal=jacsat,manuscript=article]{achemso}

\usepackage{pbox}
\author{Saeed Izadi$^\dag$}
\author{Ramu Anandakrishnan$^\ddag$}
\author{Alexey V. Onufriev$^{\P,}$}
\affiliation{
$^\dag$Department of Engineering Science and Mechanics, 
$^\ddag$Department of Computer Science, 
$^\P$Department of Computer Science and Physics, 
Virginia Tech, Blacksburg, VA 24060}
\email{alexey@cs.vt.edu}
\title%[An \textsf{achemso} demo]
  {Building Water Models, A Different Approach}

\abbreviations{OPC}
\keywords{water models, electrostatics, liquid structure simulations, optimal point charge approximation}

\begin{document}

%\begin{tocentry}
%\centering
%\parbox{.51\textwidth}{\scriptsize         
%Charge distribution of the water molecule in the gas phase obtained from a quantum mechanical calculation. %~\cite{Anandakrishnan2013}. 
%%Paradoxically, 
%Three point charges that optimally reproduce the electrostatic potential of this charge distribution are clustered in the middle, 
%as opposed to the intuitive on-nuclei placement used by common water models that results in a much poorer electrostatic description of the  charge distribution.} %underlying
%\hfill
%%\centering
%\parbox{.47\textwidth}{
%\includegraphics[width=.48\textwidth]{fig6aplos.revised.eps} %fig6plos_revised.eps
%}
%\vspace{-.02in}
%
%\end{tocentry}

\begin{abstract}
  Simplified, classical models of water are an integral part of atomistic molecular simulations,
especially in biology and chemistry where hydration effects are critical.
Yet, despite several decades of effort, these models are still far from perfect.
Presented here is an alternative approach to constructing point charge water models
-- currently, the most commonly used type. In contrast to the conventional approach,
we do not impose any geometry constraints on the model other than symmetry.
Instead, we optimize the distribution of point charges to best describe
the ``electrostatics" of the water molecule, which is key to many unusual
properties of liquid water. The search for the optimal charge distribution
is performed in 2D parameter space of key lowest multipole moments of the model,
to find best fit to a small set of bulk water properties at room temperature.
 A virtually exhaustive search is enabled via analytical equations that relate
 %[[[(the positions and values of the optimal point charges)
 %replace with
 the charge distribution to the multipole moments.
 The resulting ``optimal'' 3-charge, 4-point rigid water model (OPC) reproduces
 a comprehensive set of bulk water properties significantly more accurately than
 commonly used rigid models:
 %average and maximum errors relative to experiment are 0.76\% and 5\%, respectively.
average error relative to experiment is 0.76\%.
Close agreement
with experiment
 holds over a wide range of temperatures, well outside the ambient conditions
 at which the fit to experiment was performed. The improvements in the proposed water model extend beyond bulk properties:
 compared to the common rigid models, predicted hydration free energies of small molecules
 in OPC water are uniformly closer to experiment, root-mean-square error $< 1$ kcal/mol.

\end{abstract}

\section{Introduction}
Water is the most extensively studied molecule~\cite{Kale2012Natural,Tu2000Electronic,Dill2007Modeling}, %,Kell1975Density
yet our understanding of how this deceptively simple compound of just three 
atoms gives rise to the many extraordinary properties of its liquid 
phase~\cite{Finney2001,Finney2004,Ball1999Life} 
is far from complete~\cite{Stillinger1980}.    
The complexity of the water properties combined with multiple possible 
levels of approximations (e.g. quantum vs. classical, flexible vs. rigid) 
has led to the proposal of literally hundreds of 
theoretical and computational models for water~\cite{Guillot2002}.
Among these, the most simple and computationally efficient, 
rigid non-polarizable models that represent 
water as a set of point charges
at fixed positions relative to the oxygen nucleus stand out as 
the class used in the vast majority of biomolecular studies today.
Commonly used rigid models 
(e.g. TIP3P \cite{TIP3P} and SPC/E \cite{SPCE} 3-point models,
TIP4P/Ew \cite{TIP4PEW} 4-point model, and the TIP5P \cite{TIP5P}  
5-point model) have achieved a reasonable compromise between accuracy and speed, 
but are by no means perfect~\cite{Guillot2002,Mark2001Structure}. 
In particular, none of these models faithfully reproduce all the 
key properties of bulk water simultaneously. 
The search for more accurate yet computationally facile water models 
is still very active \cite{Wang2014pande,TIP4Peps,Wang2013iAMOEBA,Dill2012}.

\begin{figure}%[!h]
%\vspace{-.2in}
\centering
\includegraphics[width=.5\textwidth]{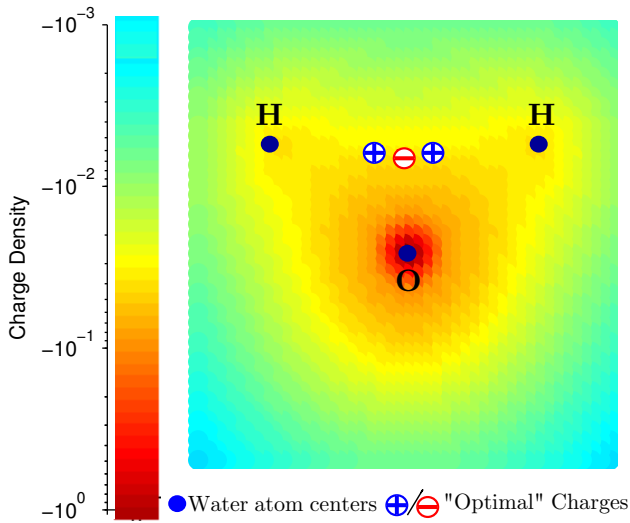} %fig6plos_revised.eps
\caption{Charge distribution of the water molecule in the gas phase obtained from a quantum mechanical calculation~\cite{Anandakrishnan2013}. 
Paradoxically, three point charges that optimally reproduce the electrostatic potential of this charge distribution are clustered in the middle, 
as opposed to the intuitive on-nuclei placement used by common water models that results in a much poorer electrostatic description of the underlying charge distribution.  }
%Optimal Point Charge distribution for water in the gas phase. The three point charges placed at (0,-0.16, 0.49), (0, 0.16, 0.49), and (0, 0, 0.47), with charge values 13e, 13e, and -26e respectively, are obtained so that the three lowest order multipole moments of the charge density are best reproduced. 
%A commonly used alternative is to place the point charges on the nuclei and set the charge values (0.32e, 0.32e, -0.64e) to match the dipole moment of the charge density.}
\label{fig:gasphase}
\end{figure}

Many unique properties of liquid water are due to the ability of the water molecules 
to establish a hydrogen-bonded structure,  
through the attraction between the electropositive hydrogen atoms 
and the electronegative oxygen atoms~\cite{Marechal2007hbond}. 
Therefore, a key challenge in developing classical water models is to find  
 an accurate yet simplified description of the charge distribution of 
the water molecule that can adequately account for the hydrogen bonding in the liquid phase. 
Procedures employed to develop commonly used rigid water models generally 
impose constraints on the geometry 
(OH bond length and HOH angle)
based on experimental observations, %data point,
most commonly by fixing the positive point charges at the hydrogen nuclei positions.
The atomic partial charges and the Lennard-Jones potential parameters 
are then optimized to reproduce selected bulk properties of water \cite{Guillot2002}.
This approach may not necessarily accurately reproduce the electrostatic
characteristics of the water molecule due to severe constraints on allowed variations in the charge distribution being optimized. %liquid water.  
The configuration of three point charges to best describe the charge distribution 
of the water molecule can be very different from 
what one may intuitively expect based on its well-known atomic structure.
Consider, for example, the gas-phase quantum-mechanical (QM) charge distribution of a
water molecule, Figure \ref{fig:gasphase}.
The shown tight clustering of the point charges away from the nuclei
reproduces the electrostatic potential around the QM charge distribution
considerably more accurately than the more traditional distribution  
with point charges placed on or near the nuclei.  
For the optimal charge placement, Figure \ref{fig:gasphase}, 
the maximum error in electrostatic potential 
at the experimental oxygen-Na$^{+}$ distance (2.23 \AA\ ) from the origin, %(RMSE=1.2 kcal/mol with maximum error of 1.4 kcal/mol), 
is almost 5.4 times smaller than that of the nucleus-centered alternative (1.4 kcal/mol vs. 7.56 kcal/mol). %(RMSE=5.3 kcal/mol with maximum error of 7.56 kcal/mol) 
%\cite{Anandakrishnan2013}. 

Intrigued by the idea that optimal placement of the point charges in a water 
model 
can be very different from the ``intuitive'' placement 
on the nuclei, and encouraged by the significant improvement of the accuracy 
of electrostatics brought about by this strategy in gas-phase, 
we explore what the approach can offer for building classical 
water models in the liquid phase. In what follows, we describe the construction
and testing of a 4-point, rigid ``optimal'' point charge (OPC)
water model. 

\section{Approach}
\begin{figure}[!h]
% \vspace*{-0.1in}
%\centering
%\includegraphics[width=0.5\linewidth]{geo_param.eps} \\ %\subfloat[][]
%\includegraphics[width=0.5\linewidth]{OPCGEOpdf.eps}  
\includegraphics[width=0.35\textwidth]{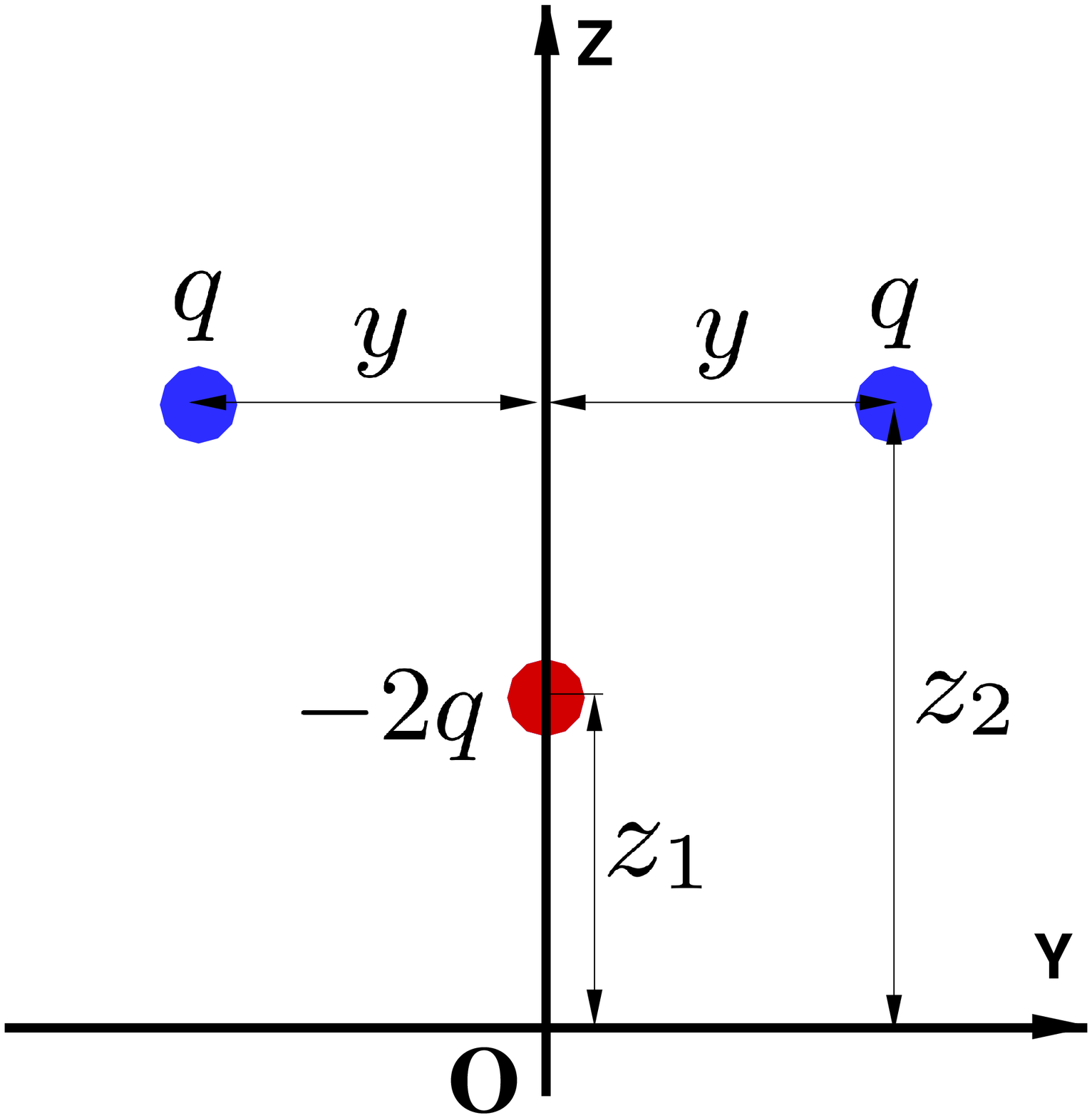} %fig6plos_revised.eps
\includegraphics[width=0.45\textwidth]{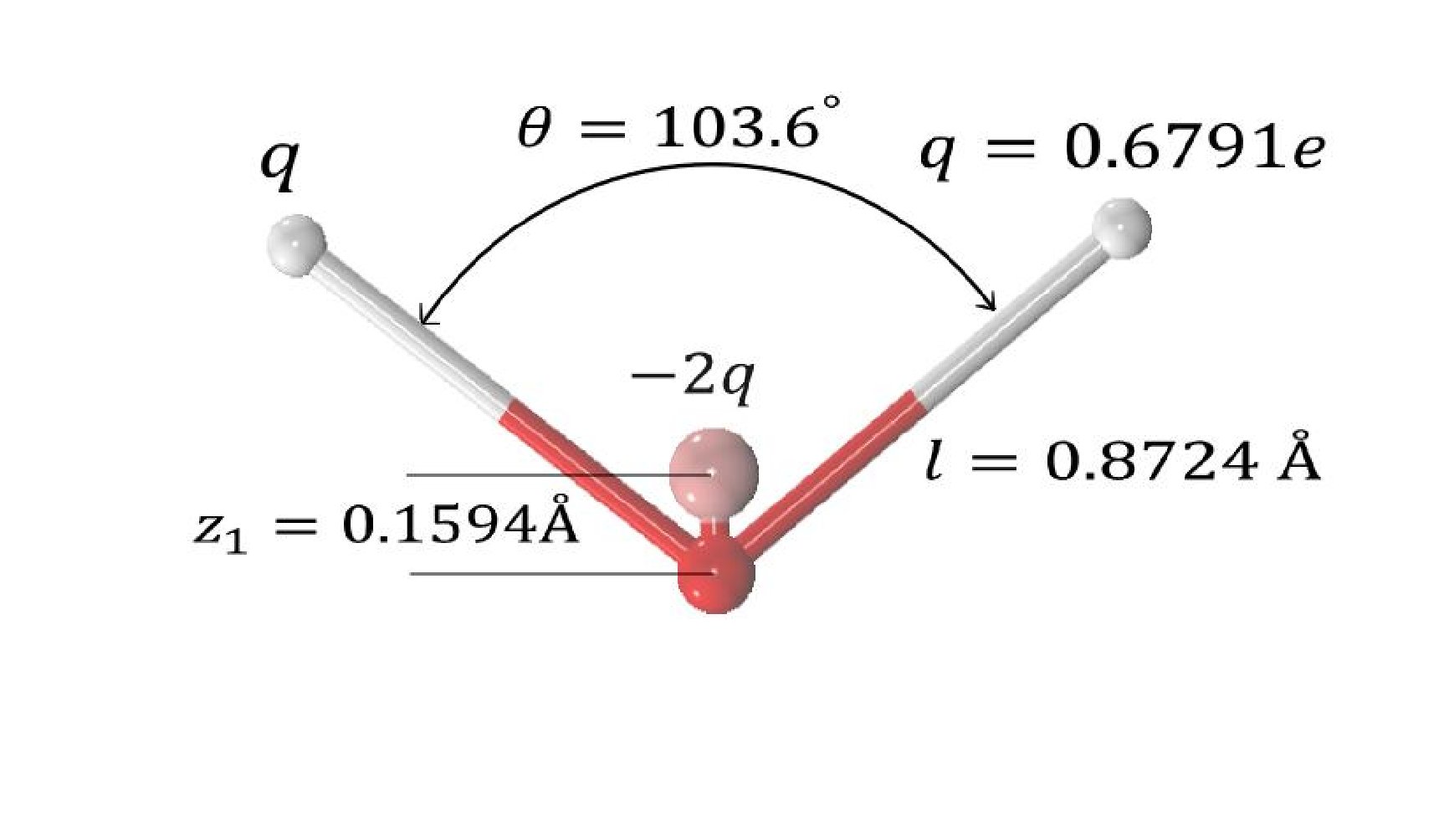} %fig6plos_revised.eps
% \vspace*{-0.15in}
\caption{\textit{\textbf{Left.}} The most general configuration for a three point charge water model consistent with $C_{2v}$ symmetry of the water molecule. 
%The charge distribution parameters ($y$, $z_{1}$, $z_{2}$, and $q$) are calculated to optimally reproduce 
%a given set of dipole, quadrupole and octupole moments.    
%The value of the positive and negative charges are q and -2q, respectively.
%The geomrty is determined by fixing y, z1 and z2. 
The single Lennard-Jones interaction is centered on the origin (oxygen). 
\textit{\textbf{Right.}} The final, optimized geometry of 
the proposed 3-charge, 4-point OPC water model. }\label{geoparamfig}
\end{figure}

Most unique properties of liquid water are due to the complexity of 
the hydrogen bonding interactions, 
which are primarily described 
by the electrostatic interactions~\cite{Morokuma1977} %,Chen1996}  %Rablen1998,
within classical potential functions, including those used in common water 
models. 
While the electrostatic interactions are complemented 
by a Lennard-Jones (LJ) potential, 
the latter is generally represented by a single site centered on the oxygen -- 
the corresponding interaction is isotropic and featureless, in contrast 
to hydrogen bonding which  is directional.
Therefore, an accurate representation of electrostatic 
interactions is paramount 
for accurately accounting for hydrogen bonding and the 
properties of liquid water. 
In a search for the best ``electrostatics'', 
commonly used distance and angle constraints 
on the configuration of a model's point charges 
are therefore of little relevance to classical rigid water models, 
yet these constraints impede the search for the ``best'' model geometry. %for the best ``electrostatics''. 
%which is of key importance for reproducing most unique features of liquid water.  
This observation leads to one of the key features of our approach: 
any ``intuitive" constraints on point charges or their geometry (other than the fundamental C$_{2v}$ symmetry of water molecule) are completely abandoned 
here in favor of 
finding an optimal electrostatic charge distribution that best approximates 
liquid properties of water.    
	While ultimately it is the values of the point charges and their 
relative positions that we seek,~(Figure~\ref{geoparamfig}), 
we argue that the conventional 
``charge--distances--angles" space~\cite{TIP3P,TIP4PEW,TIP5P,SPCE} %[[[AO: Cite lots of refs. essentially all the key water model papers]]] 
is not optimal to perform the search for the
best electrostatics model. These coordinates affect the resulting 
electrostatic potential in a convoluted manner, it is unclear which ones are 
key. On the other hand, any complex charge distribution 
can be systematically described by its multipole moments, with lower 
moments expected to have a more profound effect on liquid water 
properties~\cite{stone1997}. 
Therefore our second key proposal is to search for the optimal model 
geometry and point charges in a subspace of water
multipole moments, which we can systematically vary. 
Clearly, any reasonable water model needs to account for the  
large dipole moment of water molecule in order to 
reproduce dielectric properties of the liquid state~\cite{Dill2012,Niu2011Large}.
At short distances where hydrogen bonds between 
water molecules form ($\approx$2.8\AA), 
the relevance of higher electrostatic moments is also significant. 
For instance, the larger component of the water quadrupole
has a strong effect on the liquid water structure 
seen in simulations~\cite{Niu2011Large}, 
and on the phase diagram~\cite{Abascal2007}.  
The next order terms -- octupole moments --  
while presumably less influential, also affect water 
structure {\it e.g.} around ions~\cite{Te2010}. 
An intricate interplay between the dipole, 
quadrupole and octupole moments gives rise
to the experimentally observed charge hydration asymmetry 
of aqueous solvation --
strong dependence of hydration free energy on the sign of the solute charge 
\cite{Mukhopadhyay2012Charge,Mukhopadhyay2014}. 
Therefore, we seek a fixed-charge rigid model that optimally represents 
the three lowest order 
multipole moments of the water molecule. %, in a manner described below. 
The exhaustive search for the optimum is enabled by 
the third key feature of our approach: a set of analytical equations 
that relates key multipole moments to the positions and values of the 
point charges of the water model. 
%A search for the best fit in the space of key multipole moments is the centerpiece of our approach. 

\paragraph{The specifics.} 
To optimally reproduce the three lowest order multipole moments for 
the water molecule charge distribution, 
a minimum of three point charges are needed~\cite{Anandakrishnan2013}. 
The most general configuration for a three point charge model 
consistent with $C_{2v}$ symmetry of the water molecule is shown in 
Figure \ref{geoparamfig}: the 
point charges are placed in a V-shaped pattern in the Y-Z plane. We follow 
convention~\cite{TIP3P,TIP4PEW,TIP5P,SPCE} %[[[AO: cite the same lot of models ]]] 
and place the single 
Lennard-Jones (LJ) site on the oxygen atom. 
The four parameters ($q$,$z_{2}$,$z_{1}$ and $y$) that completely define the charge distribution, (Figure~\ref{geoparamfig}),  
are uniquely determined via analytical equations introduced in~\emph{Methods}, % (Eqs.~\ref{OPCsolutions1},\ref{OPCsolutions2} and SI), 
to best reproduce a targeted set of three lowest order 
multipole moments (dipole, quadrupole and octupole)~\cite{Anandakrishnan2013} 
as detailed below.  
%which greatly simplify and speed up the search. 
The ability to independently vary the moments of the charge distribution, 
provided by these analytical expressions, 
allows a full exploration in the relevant subspace of the moments.
Generally, the importance of the multipole moments 
are inversely related to their order. 
The highest order multipole moment here 
is the octupole that has two independent components ($\Omega_{0}$ and $\Omega_{T}$), which we fix  
%Therefore, we fix the two components of the octupole moment 
to high quality quantum mechanical (QM) predictions, QM/230TIP5P \cite{QM230TIP5P}, Table \ref{table:momentstable}. 
The linear component of the quadrupole $Q_{0}$ is known to be relatively  
small for the water molecule and 
not expected to be very important~\cite{Rick2004TIP5PEW}, 
therefore, we also simply set it to the known QM value ( 
QM/230TIP5P \cite{QM230TIP5P}, Table \ref{table:momentstable} ). 
%The remaining key multipole moments ($Q_{0},\Omega_{T},\Omega_{0}$) were set to the values from QM/230TIP5P \cite{QM230TIP5P} (see Table \ref{table:momentstable}).
This leaves the two most important components, 
the dipole ($\mu$) and the square quadrupole ($Q_{T}$), as the two 
key search parameters we vary.  
We attempt to find the best fit to six key bulk properties by exhaustively searching 
in the 2D space of $\mu$ and $Q_{T}$, Figure~\ref{scores}, 
within the ranges that reflect 
known experimental uncertainties~\cite{Gregory1997expdipole} and those of QM calculations~\cite{AIMD2,AIMD1}, Table~\ref{table:momentstable}. The six 
target bulk properties are: static dielectric constant $\epsilon_{0}$,
 self diffusion coefficient $D$,  heat of vaporization $\Delta H_{vap}$,
 density $\rho$ and the position $roo1$ and height $g(roo1)$ 
of the first peak in
oxygen-oxygen pair distribution functions.
These properties are calculated from molecular dynamics (MD) simulations,
see \emph{Methods} and the SI.
For every trial value of $\mu$ and $Q_{T}$ (and the fixed values of $Q_{0}$, $\Omega_{0}$ and $\Omega_{T}$), 
 %we use a set of analytical equations, see \emph{Methods}, to determine the charge distribution parameters ($q$,$z_{2}$,$z_{1}$ and $y$).  
 the charge distribution parameters ($q$,$z_{2}$,$z_{1}$ and $y$) are analytically determined (see \emph{Methods}).

 \begin{table}[!h]
% \centering
\caption{Water molecule multipole moments centered on oxygen: from experiment, common rigid models, 
   liquid phase quantum calculations, and OPC model (this work).} \label{table:momentstable}
\begin{tabular}{lccccc} %{@{\vrule height 3.0pt depth1pt  width0pt}lccccc}
\hline
  & $\mu$ & $Q_{0}$ & $Q_{T}$& $\Omega_{0}$ & $\Omega_{T}$\\
 Model & [D] & [D\AA] & [D\AA] & [D\AA$^{2}$] & [D\AA$^{2}$]\\
\hline
EXP~(liquid)~\cite{Gregory1997expdipole} &  2.5$-$3 & NA &  NA & NA & NA \\  %Clough1973Dipole
SPC/E & 2.35 & 0.00 & 2.04 & -1.57 & 1.96 \\
TIP3P & 2.35 & 0.23 & 1.72 & -1.21 & 1.68 \\
TIP4P/Ew  & 2.32 & 0.21 & 2.16 & -1.53 & 2.11 \\
TIP5P & 2.29 & 0.13 & 1.56 & -1.01 & 0.59 \\
AIMD1~\cite{AIMD1}   & 2.95 & 0.18 & 3.27 & NA & NA \\
AIMD2~\cite{AIMD2}   & 2.43 & 0.10 & 2.72 & NA & NA \\
QM/4MM~\cite{Niu2011Large}  & 2.49 & 0.13 & 2.93 & -1.73 & 2.09 \\
QM/4TIP5P~\cite{Niu2011Large} & 2.69 & 0.26 & 2.95 & -1.70 & 2.08 \\
QM/230TIP5P~\cite{QM230TIP5P} & 2.55 & 0.20 & 2.81 & -1.52 & 2.05\\
\textbf{OPC} & \textbf{2.48} & \textbf{0.20} & \textbf{2.3} & \textbf{-1.484} & \textbf{2.068}
 \\ 
\hline
\end{tabular}
\end{table}

For every charge distribution calculated as above, 
%The LJ site is centered on the oxygen and is characterized by ($A_{LJ}$ and $B_{LJ}$) parameters. 
%are determined following 
%the standard procedure in the literature \cite{Rick2004TIP5PEW}. 
%For each value of $\mu$ and $Q_{T}$, 
%we independently determine the best values of  Lennard-Jones parameters 
%($A_{LJ}$ and $B_{LJ}$) as follows. 
the value $A_{LJ}$ of the 12-6 Lennard-Jones (LJ) potential, 
which is mainly responsible for the liquid structure~\cite{Rick2004TIP5PEW}, %[[[AO Do we have a citation for this statement? ]]]  
%[[[SI: which is responsible for short-ranged repulsive interactions~\cite{Rick2004TIP5PEW}]]]
is selected so that the location of the first peak $g_{oo}(r)$ 
of the oxygen-oxygen radial distribution function (RDF)
is in agreement with recent experiment~\cite{Skinner2013} (see \emph{Methods}). %[[[AO: Are tolerances described in "Methods"? If so, say "see Methods" ]]] 
The value of $B_{LJ}$ is optimized so that the experimental 
value for density is achieved.  
The parameters $A_{LJ}$ and $B_{LJ}$ can be optimized nearly 
independently due to the weak coupling between them~\cite{Rick2004TIP5PEW}. %[[[AO: See SI? ]]] 
%We find that $A_{LJ}$ and $B_{LJ}$ can be optimized nearly 
%independently due to the weak coupling between them~\cite{Rick2004TIP5PEW}. %[[[AO: See SI? ]]]  
%$A_{LJ}$ and $B_{LJ}$ parameters provides a suitable 
%procedure for automated parameter optimization. 
%Because the LJ potential site may or may not be coincident with 
%the negative charge calculated above, 
%the resulting model will include four points overall.

%For every charge distribution calculated as above, 
%the usual 12-6 Lennard-Jones (LJ) potential is employed to model the van der Waals interaction among the oxygens. 
%The 12-6 Lennard-Jones parameters ($A_{LJ}$ and $B_{LJ}$) are determined following 
%the standard procedure in the literature \cite{Rick2004TIP5PEW}. 
%%For each value of $\mu$ and $Q_{T}$, 
%%we independently determine the best values of  Lennard-Jones parameters 
%%($A_{LJ}$ and $B_{LJ}$) as follows. 
%The value of $A_{LJ}$, which is mainly responsible for characterizing the liquid structure, 
%is selected so that the location of the first peak $g_{oo}(r)$ 
%of the oxygen-oxygen radial distribution function (RDF)
%is in agreement with experiment \cite{Skinner2013}.
%The value of $B_{LJ}$ is optimized so that the experimental 
%value for density is achieved.  
%The weak coupling of $A_{LJ}$ and $B_{LJ}$ parameters provides a suitable 
%procedure for automated parameter optimization. 
%Because the LJ potential site may or may not be coincident with 
%the negative charge calculated above, 
%the resulting model will include four points overall.
The result of the above search procedure is a ``quality map" of all possible 
water models in the $\mu-Q_{T}$ space: 
the proposed OPC model is the one with the highest quality score. 
 
%The fine-grain search for the best fit in the $\mu-Q_{T}$ space is the 
%centerpiece of our approach. 
%The quality score of each tested water model characterized by 
%a given set of $\mu$ and $Q_{T}$ is assessed 
%by comparing the computed values for a set of key bulk properties 
%to their experimental values.   
%The final OPC model is the one with the highest quality score 
%in the $\mu-Q_{T}$ space. 

\section{Results and discussion}
\subsection{The proposed optimal point charge model}

\begin{figure}[!ht]
%\vspace{-.15in}
\centerline{\includegraphics[width=.8\linewidth]{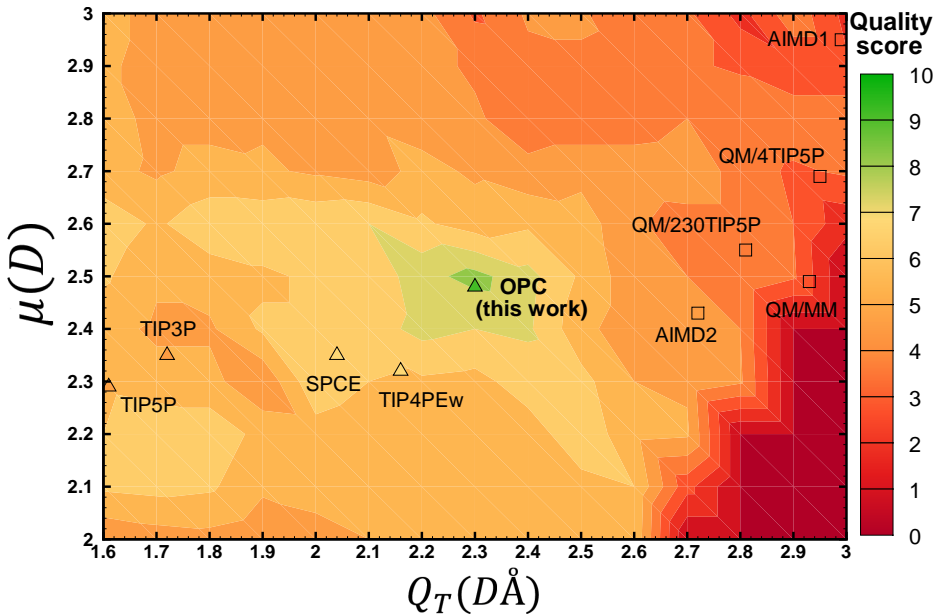}}
%\centerline{\includegraphics[width=1.1\linewidth]{mu-QTplot_final-vm.eps}}
\caption{The quality score distribution of test water models in the space of dipole ($\mu$) and quadrupole ($Q_{T}$).   
Scores (from 0 to 10) are calculated based on the accuracy of predicted 
values for six key properties of liquid water (see text). 
The resulting proposed optimal model is termed OPC. 
For reference, the $\mu$ and $Q_{T}$ values of several commonly used water 
models (triangles, quality score given by the color at the symbol position) 
and quantum calculations (squares) are placed on 
the same map (see also Table~\ref{table:momentstable}).  
The actual positions of AIMD1 and TIP5P are slightly modified to fit 
in the range shown.  
%The red region in bottom right corresponds to degenerate cases where the point charges converge to a singular point.
}\label{scores}
\end{figure}

	As described above, we have performed 
an exhaustive search in the $\mu-Q_{T}$ space 
for the best fit to six target bulk properties of 
liquid water at ambient conditions, Figure \ref{scores}. 
The entire region of the $\mu-Q_{T}$ space was mapped out using initially a 
relatively coarse grid spacing (0.1~D and 0.1~D\AA) 
in each direction shown in Figure \ref{scores}. 
%The remaining key multipole moments ($Q_{0},\Omega_{T},\Omega_{0}$) were set to the values from QM/230TIP5P \cite{QM230TIP5P} (see Table \ref{table:momentstable}).        
At this point, 
the quality of each test water model -- corresponding to a $\mu, Q_{T}$ point  
 on the map -- 
is characterized by a quality score function (see~\emph{Methods}) 
from a recent 
comprehensive review~\cite{Vega2011} %[[[AO: cite that paper. Vega? ]]],
 based on the same six key bulk properties used for the fitting. 

 Accordingly each model is assigned a quality score, using the score function explained in the \emph{Methods} section, and is shown in Figure \ref{scores}. 
As demonstrated in Figure~\ref{scores}, the highest quality region (the green area) occurs for (2.4~D~$\leq$~$\mu$~$\leq$~2.6~D) and (2.2~D\AA$~\leq~Q_{T}~\leq$~2.4~D\AA). 
 %It is interesting to notice that this region is in accordance with the predictions for the dipole, and the moment ratio ($\mu/Q_{T}~1.1$).
The region is relatively small and this is why an exhaustive, 
fine-grain search was required to identify the best model, 
which we refer to as the Optimal Point Charge (OPC) model (Figure \ref{scores}). 
 
From Figure~\ref{scores}, one can see three distinct regions in the $\mu-Q_{T}$ space: 
the ``common water models" region with relatively small dipole and 
square quadrupole moments, 
the ``QM" region characterized by larger dipole and square quadrupole, 
and narrow, high quality (OPC) region with intermediate values of these
two key moments.         
%The OPC model is clearly departed from the region of common models and 
%is the most close to the QM region.
Compared to the other rigid models shown, OPC reproduces 
the multipole moments of water molecule in the liquid phase substantially better. 
In fact, the OPC dipole moment (2.48~D) is in best agreement with the values 
from QM calculations and experiment.  
OPC's $Q_{T}$ (2.3~D\AA) is larger than the corresponding 
values of the common models,  
and is closest to the QM predictions (Figure~\ref{scores}, Table~\ref{table:momentstable}). 
By construction, OPC's small $Q_{0}$ component of the quadrupole matches
the reference QM value, % the selected QM value, 
and its octupole moments are the best approximations.  
%(for a given $\mu$ and $Q_{T}$) within the 3 point charge models (Table \ref{table:momentstable}).   
The improved accuracy of the OPC moments %with respect to commonly used models  
is an immediate consequence of the focus on electrostatics and the 
unrestricted fine-grain search in the $\mu-Q_{T}$ space, which we believe 
is the most relevant subspace of the water multipole moments at this level of 
approximation.  
This important improvement became possible through the abandoning of the conventional  
geometrical constraints, 
allowing the moments to be varied independently; 
the availability of analytical equations 
that connected the charge distributions with multipole moments played 
an important role too. 

While the OPC moments are closest to the QM values, 
%which we consider as a positive development, 
they (in particular $Q_{T}$) still deviate from the QM predictions (Table~\ref{table:momentstable}, Figure~\ref{scores}).   
The low quality of the test models in which the moments were close 
to the QM values (squares, Figure~\ref{scores})
suggests that, within the 3-charge models explored here, an analytical fit 
of moments to QM predictions does not 
guarantee agreement with experimental liquid phase properties. 
This discrepancy can be due to a number of limitations and approximations
inherent to classical, rigid, non-polarizable water models, see {e.g.}~Refs.~\cite{Vega2011,Guillot2002,Wang2013iAMOEBA}. 
Based on our own results, we suggest that another important factor may be 
the small number of point charges used to represent the complex 
charge distribution of real water molecule.   
Namely, a three point charge model is fundamentally unable to exactly reproduce 
the reference dipole, quadrupole and octupole moments simultaneously~\cite{Anandakrishnan2013}, %, [[[AO: Reference the OPCA paper? ]]] 
and essentially has no control over the accuracy of its moments beyond the octupole. 
The contribution of the higher order multipole moments 
to electrostatic potential 
%may be relatively small far from the charge distribution, their contribution 
can be significant at close distances, which are 
relevant to water-water and water-ion interactions in liquid phase.   
We conjecture that the relatively small $\mu$ and $Q_{T}$ value found at 
the highest quality region (green zone, Figure~\ref{scores})
compared to QM predictions (squares, Figure~\ref{scores}),  
may be a compromise to keep the  higher moments not too far from optimal, 
ensuring a reasonable net electrostatic potential.

\begin{table} [t]
%\vspace*{-.1in}
\caption{Force field parameters of OPC and some common rigid models. Units for $A_{LJ}$ and $B_{LJ}$ are ($10^{3}$\AA$^{12}$kcal)/mol and (\AA$^{6}$kcal)/mol, respectively. For comparison, water molecule geometry in the gas phase is also included.}  \label{table:forcefield} %
\begin{tabular}{lcccccc} %{*{1}{p{1.2cm}}*{6}{p{.8cm}}}
\hline
%{@{\vrule height 5.5pt depth3pt  width0pt}lrrcccc}
%
%&\multicolumn5c{Repeat length}\\
%\noalign{\vskip-11pt}
%Age of onset,\\
%\cline{2-6}
& $q[e]$ &  $\l$[\AA] & $z_{1}$[\AA] & $\Theta[deg]$ & $A_{LJ}$ & $B_{LJ}$\\
\hline
  EXP(gas) & NA & 0.9572 & NA & 104.52 & NA & NA \\ %  
     TIP3P & 0.417 & 0.9572 & NA & 104.52 & 582.0 & 595.0  \\ % inserting 3.16435 & 0.16275
     TIP4PEw & 0.5242 & 0.9572 & 0.125 & 104.52 & 656.1 & 653.5  \\ % inserting 3.16435 & 0.16275
     TIP5P & 0.241 & 0.9572 & NA & 104.52 & 544.5 & 590.3  \\ % inserting 3.16435 & 0.16275
     SPC/E & 0.4238 & 1.0 & NA & 109.47 & 629.4 & 625.5 \\ % inserting
     \textbf{OPC} & \textbf{0.6791} & \textbf{0.8724} & \textbf{0.1594} & \textbf{103.6} & \textbf{865.1} & \textbf{858.1}  \\ % inserting body of the \textbf{3.1665} &
\hline
\end{tabular}
\end{table}

The OPC point charge positions and values and the LJ parameters are listed in Table~\ref{table:forcefield}.        
The $|O-q^{+}|$ distances for OPC are shorter (0.8724\AA), 
and the $\angle~q^{+}Oq^{+}$ angle (Figure~\ref{geoparamfig}) is slightly narrower ($103.6^{\circ}$) 
than the corresponding experimental values of $|O-H|$ bond and $\angle$HOH angle %[[[AO: there is a way in later to add an overhat for angle. Don't worry if cn't find ]]]  
for the water molecule in the gas phase (0.9572\AA ~and $104.52^{\circ}$). 
The charge magnitudes of the OPC model are significantly 
larger than those of other common models (Table~\ref{table:forcefield}). 
Although the OPC charge distribution is not as tightly clustered as the configuration of the optimal charge 
model in the gas phase (Figure~\ref{fig:gasphase}), %with respect to experimental geometry of the water molecule. 
the deviation of OPC geometry from that of other models and the water molecule in the gas phase is influential. 
In particular, the quality of water models is extremely sensitive to the values of electrostatic multipole moments (Figure~\ref{scores}), 
which by itself are very sensitive to the geometrical parameters (Eqs.~\ref{equdipole}-\ref{equQt}, and SI).

\subsection{Bulk properties}
Since the geometry of the proposed   
rigid, non-polarizable 
OPC model optimized for the liquid phase, Figure~\ref{geoparamfig}, 
is very different from the expected optimum 
outside of the liquid phase, Figure~\ref{fig:gasphase}, 
here we test OPC model in the liquid phase only. 
% A summary of the properties at ambient conditions for OPC is presented in Table \ref{table:bulkproperties}, 
%along with the properties for other commonly used models, and from experiment. 
The quality of the model in reproducing experimental bulk water 
properties at ambient conditions, and a comparison  
with other most commonly used rigid models is presented in Table~3. %\ref{table:bulkproperties}.  
For 11 key liquid properties (Table~3) against which water models are most often benchmarked \cite{Vega2011,Vega2009ice,TIP4PEW}, 
our proposed model is within 1.8$\%$ of the corresponding experimental value, 
except for one property (thermal expansion coefficient) that deviates from 
experiment by about 5$\%$. The full oxygen-oxygen radial distribution 
function (RDF), $g(roo)$, is presented in the SI. %\ref{rdfequ}. 
%The X-ray scattering data for the liquid structure %is extracted from a recent study \cite{Skinner2013}. 
By design, the experimental position of first peak in RDF is accurately reproduced by OPC. 
The position and height of other peaks are also closely reproduced. 
%A comparison of OPC's RDF function with that of other water models is presented in the SI.  
    
\begin{table}[t]
\caption{Model vs. experimental bulk properties of water at ambient conditions (298.16 K, 1 bar): dipole $\mu$, 
density $\rho$, static dielectric constant $\epsilon_{0}$, 
self diffusion coefficient $D$, heat of vaporization $\Delta H_{vap}$, 
first peak position in the RDF $roo1$, %coordination number $n_{c}$, 
propensity for charge hydration asymmetry (CHA)~\cite{Mukhopadhyay2012Charge,Mobley2008,Rajamani2004}, isobaric heat capacity $C_{p}$, 
thermal expansion coefficient $\alpha_{p}$, and isothermal compressibility $\kappa_{T}$. 
The temperature of maximum density (TMD) is also shown. 
Bold fonts denote the values that are closest to the corresponding 
experimental data (EXP). 
Statistical uncertainties ($\pm$) are given where appropriate.} \label{table:bulkproperties} %
\begin{tabular}{*{1}{p{3.3cm}}*{4}{p{1.8cm}}*{1}{p{2.2cm}}*{1}{p{1.4cm}}} %{lcccccc} %{\hsize}
\hline
Property&TIP4PEw~\cite{TIP4PEW}& SPCE~\cite{Vega2011,Wang2014pande}&TIP3P~\cite{Vega2011,TIP5P}&
TIP5P~\cite{Vega2011,TIP5P}&\textbf{OPC}&
EXP~\cite{Vega2011,Vega2009ice,Skinner2013}\\

%\multicolumn1c{Property}&\multicolumn1c{TIP4PEw~\cite{TIP4PEW}}&
%%{{\tiny}} & {{\tiny\cite{Vega2011,Wang2014pande}}} & {{\tiny\cite{Vega2011,TIP5P}}} & {{\tiny\cite{Vega2011,TIP5P}}} & & \tiny\cite{Vega2011,Vega2009ice}}
%\multicolumn1c{SPCE~\cite{Vega2011,Wang2014pande}}&\multicolumn1c{TIP3P~\cite{Vega2011,TIP5P}}&
%\multicolumn1c{TIP5P~\cite{Vega2011,TIP5P}}&\multicolumn1c{\textbf{OPC}}&
%\multicolumn1c{EXP~\cite{Vega2011,Vega2009ice,Skinner2013}}\cr
\hline
    $\mu$($D$)             & 2.32 & 2.352 & 2.348 & 2.29 & \textbf{2.48} & 2.5--3 \\ 
%    $Q_{T}$($D\AA$) & 2.16 & 2.04 & 1.72 & 1.56 & \textbf{2.30} & - \\  
     $\rho [g/cm^{3}]$      & 0.995   & 0.994 & 0.980 & 0.979 & \textbf{0.997$\pm$0.001} & 0.997 \\        
     $\epsilon_{0}$         & 63.90      & 68   & 94    & 92    & \textbf{78.4$\pm$0.6} & 78.4 \\
     $D[10^{9}m^{2}/s]$     & 2.44   & 2.54  & 5.5  & 2.78  & \textbf{2.3$\pm$0.02} & 2.3 \\
     $\Delta H_{vap}[kcal/mol]$ & 10.58 & 10.43 & 10.26 & 10.46 & \textbf{10.57$\pm$0.004} & 10.52 \\
     %$\Delta H_{\tiny vap}[kcal/mol]$ & $10.58$ & $10.43$ & $10.26$ & $10.38$ & \textbf{10.57$\pm$0.004} & $10.52$ \\ 
     $roo1$\footnotesize[\AA]          & 2.755      & 2.75 & 2.77  & 2.75 & \textbf{2.80} & 2.80 \\
   %  $n_{c}(r_{c}=3.3$\footnotesize \AA)  & \textbf{4.31} & $4.34$ & 	$4.35$ &  	-	& $4.39$ &	$4.3$ \\ %($r_{cut}=3.30A$)
%     g(roo1) & 3.1601 & 3.050 & 2.790 & 	2.790 & 3.200 & 2.575 \\
     $CHA$~$propensity$~\textsuperscript{\emph{a}}
     & 0.52 & 0.42 & 0.43 & 0.13 & \textbf{0.51} & 0.51 \\
    $C_{p}[cal/(K.mol)]$ & 19.2 & 20.7 & 18.74 & 29 & \textbf{18.0$\pm$0.05} & 18 \\
    $\alpha_{p}[10^{-4}K^{-1}] $ & 3.2 & 5.0 & 9.2 & 6.3 & \textbf{2.7$\pm$0.1} & 2.56 \\
    $\kappa_{T}[10^{-6}bar^{-1}] $ & 48.1 & 46.1 & 57.4 & 41 & \textbf{45.5$\pm$1} & 45.3 \\
    $TMD [K]$  & 276 & 241 & 182 & \textbf{277} & 272$\pm1$ & 277 \ \ \\
\hline
\end{tabular}

   \textsuperscript{\emph{a}}  Values are calculated in this work. 
     The experimental value is a theoretical estimate \cite{Mukhopadhyay2012Charge} based on 
    experimental hydration energies of $K^{+}/F^{-}$ pair \cite{Schmid2000}. 
    See SI for details.
%\vspace*{-.25in}
\end{table}

\begin{figure}%[!hbp]  %* no [h]
%\vspace*{2.8in}
\centerline{\includegraphics[width=0.8\linewidth]{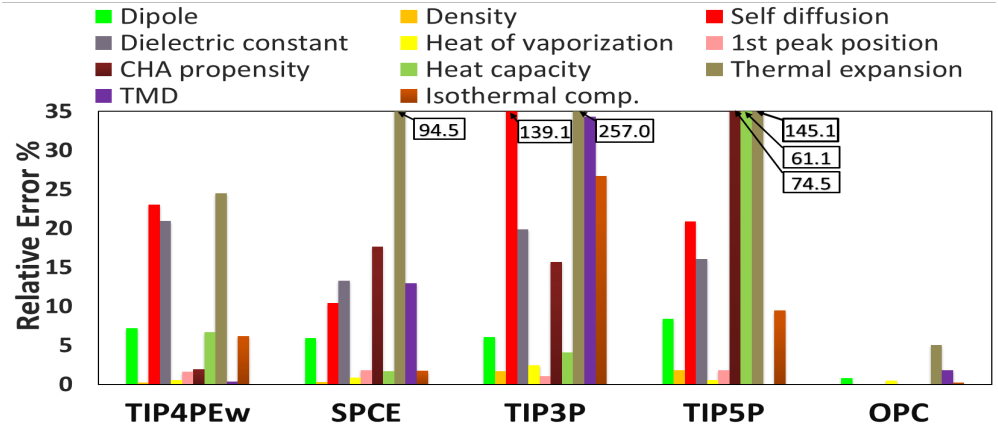}}
%\centerline{\includegraphics[width=0.8\linewidth]{barchartnofill.eps}}
\caption{Relative error in various properties by the common rigid models and OPC (this work). 
Values of the errors that are cut off at the top are given in the boxes.}\label{barchart}
\end{figure} 
While commonly used models may be in good agreement
with experiment  
for certain properties, Figure~\ref{barchart}, they often produce large errors (sometimes amounting to over 250$\%$) in some other key properties. 
In contrast, OPC shows a uniformly good agreement across all the 
bulk properties considered here.  
    
\begin{figure}[!ht]
%\FloatBarrier
%\vspace*{-.31in}
\centering
\includegraphics[width=0.4\linewidth]{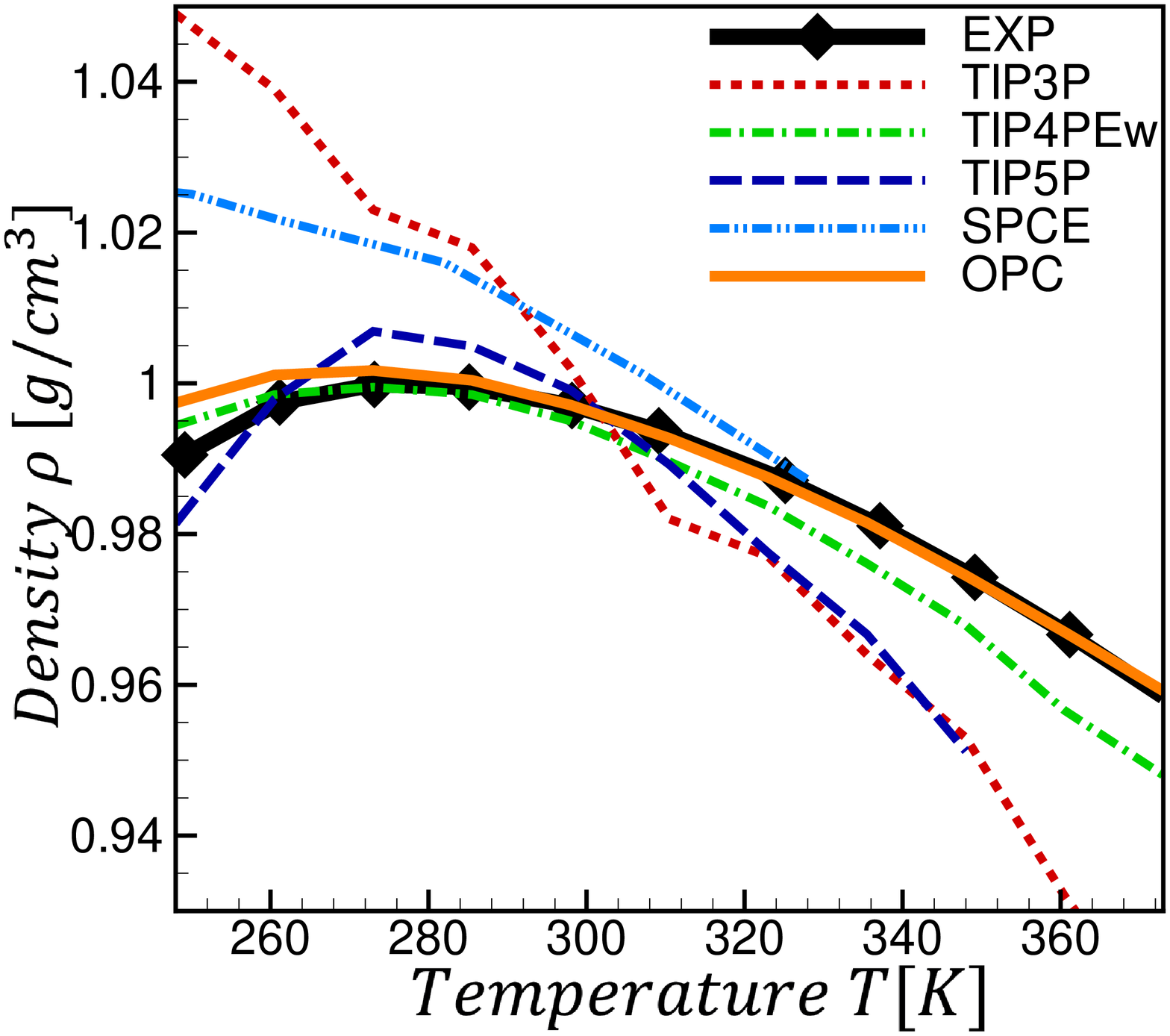} %\\ %\subfloat[][]
%(a)\\
\includegraphics[width=0.4\linewidth]{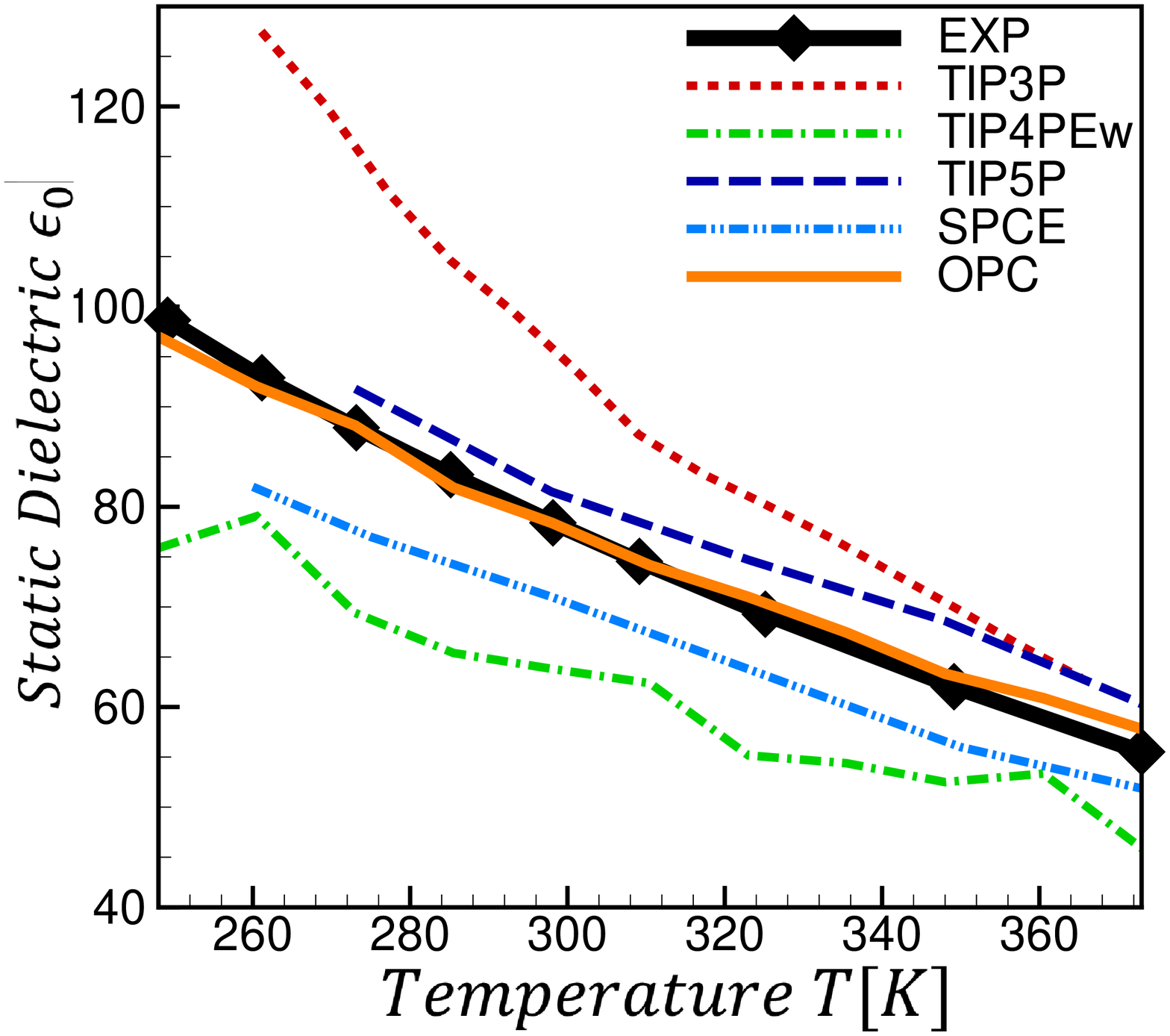} \\ 
\footnotesize ~~~~~~~~~~~~~~~~~~~~(a) Bulk density~~~~~~~~~~~~~~~~~~~~~~~~~~~(b) Static dielectric constant\\
\includegraphics[width=0.4\linewidth]{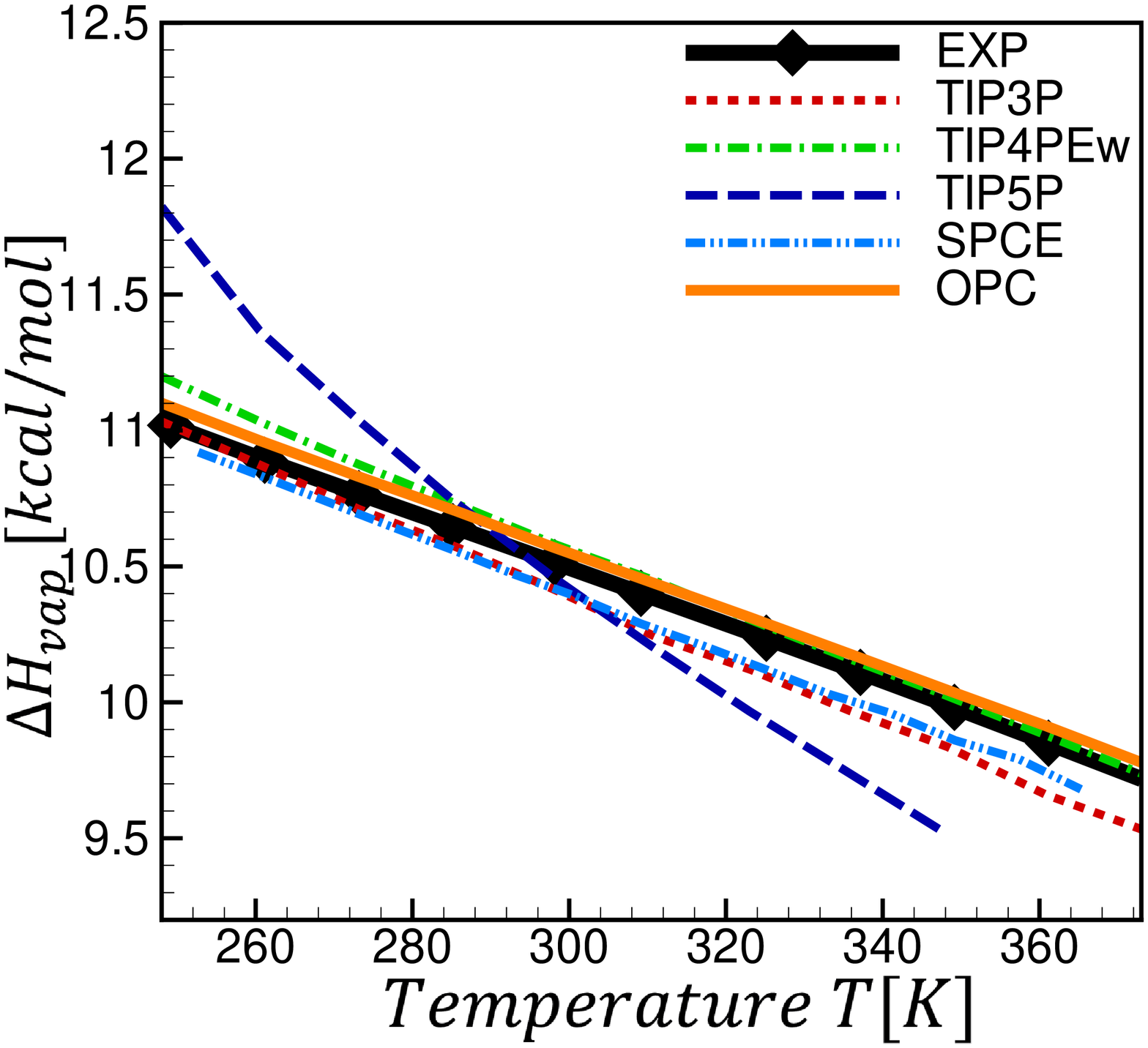} %\\
%(c)\\
\includegraphics[width=0.4\linewidth]{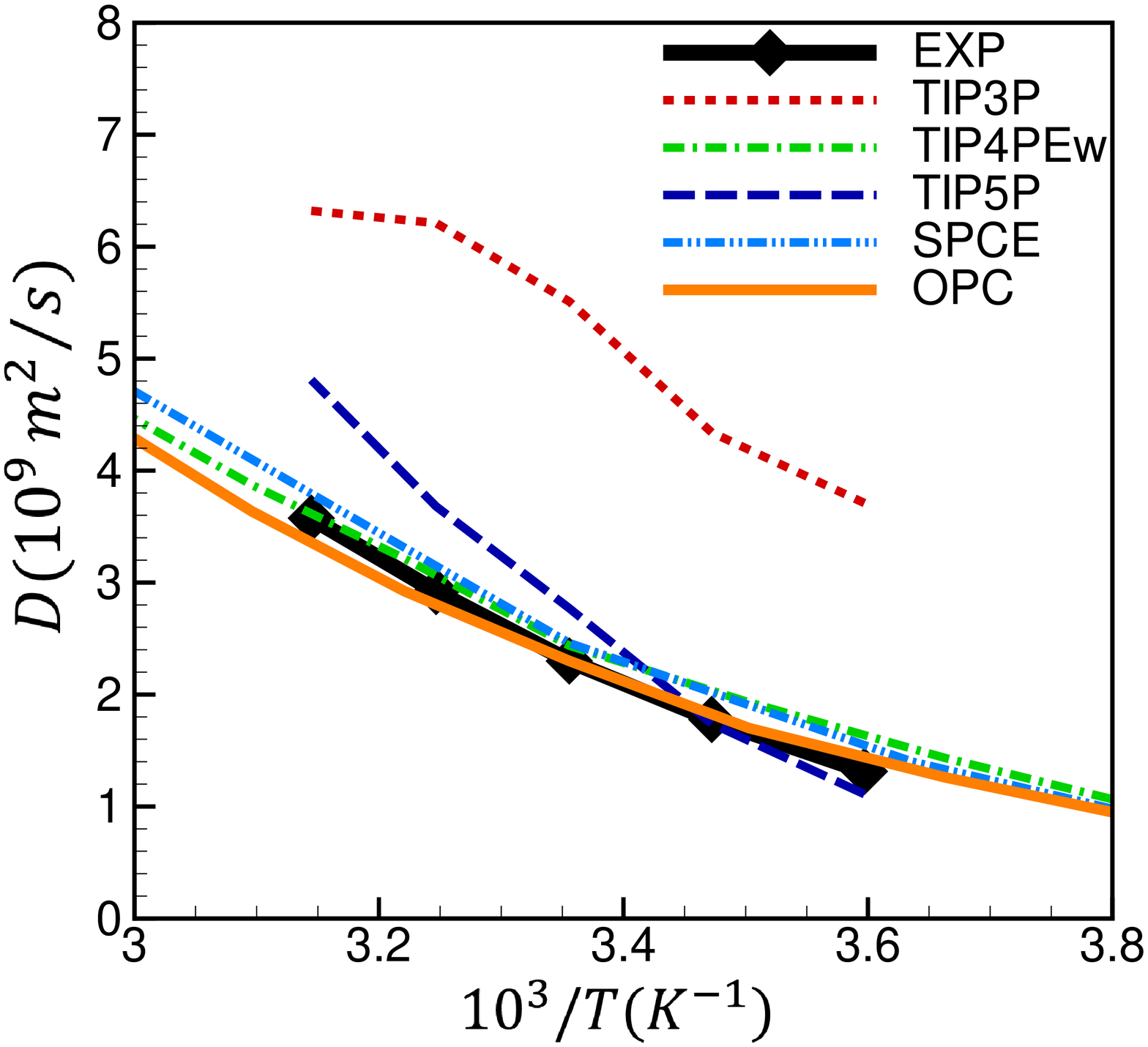}\\
\footnotesize ~~~~~~~~~~~~~~~~~~~~(c) Heat of vaporization~~~~~~~~~~~~~~~~~~(d) Self diffusion coefficient\\
\caption{ Calculated temperature dependence of water properties compared to experiment and 
several common rigid water models.
% (a) Bulk density (b) Static dielectric constant (c) Heat of vaporization (d) Self diffusion constant. 
TIP4PEw results are from \cite{TIP4PEW}, 
TIP5P from \cite{TIP5P, Vega2009ice, TIP4PEW},
%TIP3P from \cite{TIP3P, Vega2009ice, dielec_to_be_from_Wang2014pande,Jorgensen1998tip3pspctip4p},
%SPCE from \cite{English2005spcetempdep, diff_pandey,hvap_from_pande,cp_from_pandey}.
TIP3P from \cite{TIP3P, Vega2009ice, Wang2014pande,Jorgensen1998tip3pspctip4p}, 
SPCE from \cite{English2005spcetempdep,Wang2014pande}.
}\label{tempdepfig}
 
\end{figure}

The ability of OPC to reproduce the temperature dependence 
of six key water properties is shown in Figure~\ref{tempdepfig} (and SI). 
OPC is uniformly closest to experiment. 
%in all the six temperature dependent results studied here compared to all the water common models. 
It is noteworthy that the OPC model, which  
%Remarkably, the significant overall improvement in accuracy 
resulted from a search in the space of only two parameters 
($\mu$ and $Q_{T}$) 
at only one thermodynamic condition (298.16 K and 1 bar) to 
fit a small subset of bulk properties, 
automatically reproduces a large number of bulk properties 
with a high accuracy across a wide range of temperatures.   
This is in contrast not only to commonly used, but also 
to some recent rigid \cite{Wang2014pande,TIP4Peps} 
and polarizable models \cite{Wang2013iAMOEBA} 
that generally employ massive and more specialized fits against multiple properties 
over a wide range of thermodynamic conditions. 
%yet the overall end result is not more accurate than OPC (see SI). 
While noticeable advance in the accuracy of bulk properties is made by 
these latest models, 
the overall end result is not more accurate than OPC (see SI).

% Figure \ref{tempdepfig} compares the temperature dependent data from OPC, other water models and experiments, 
%for density, dielectric constant, self diffusion coefficient and heat of vaporization. 
%The temperature dependent results for heat capacity and isothermal compressibility are presented  
%in Figure \ref{tempdepcp}.   
%The OPC model was only parametrized at the ambient condition 
%but reproduces the bulk properties of liquid
%with a high accuracy across a wide range of temperatures.  
%The bulk-density $\rho$ as a function of temperature and 1 bar are shown in Figure \ref{tempdepfig}(a).
 
%While OPC slightly overestimates the density at the super cooled temperatures, 
%excellent agreement is achieved at higher temperatures. 
%The temperature of maximum density (TMD) is 272 K for the OPC model,
%which is slightly different from the experimental value (277 K). 
%TIP5P and TIP4PEw give more accurate approximations for TMD, but 
%their temperature dependence is overall less accurate than OPC. 
 
\vspace{-.1in}

\subsection{Beyond bulk properties, OPC improvements matter for practical calculations}

\begin{figure}%[t]  %* no [h]
%\vspace*{2.7in}
\centerline{\includegraphics[width=.9\linewidth]{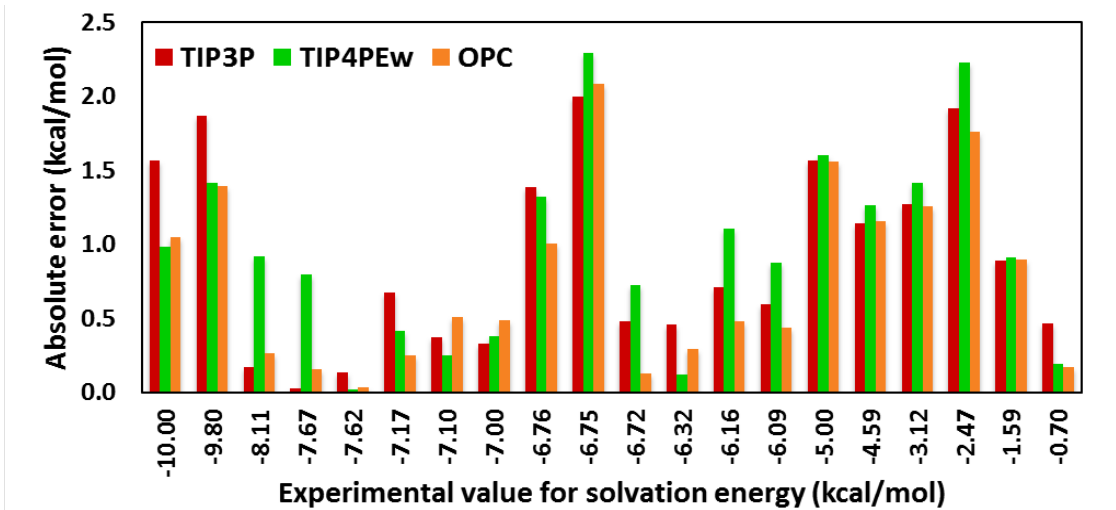}}
%\centerline{\includegraphics[width=.9\linewidth]{solvationnofill.eps}}
\caption{Absolute error in solvation free energies of a set of 20 small molecules calculated using TIP3P, TIP4P-Ew and the proposed OPC models. 
%The simulation protocol for these calculations is presented in \emph{Methods} section.
}\label{solvation}
%\vspace*{-.1in}
\end{figure}

%In the previous sections it was shown that improved electrostatics 
%yields a significantly increased accuracy in the calculation 
%of water bulk properties. 
One of the main goals of developing better water models 
is improving the accuracy of simulated 
hydration effects in molecular systems.  
Here we show that the optimized charge distribution of OPC model 
does lead to a more accurate representation 
of solute-water interactions, whose accuracy is critical to the outcomes 
of atomistic simulations.
One of the most sensitive measure of 
the balance of intermolecular and 
solute-water interaction is hydration free energy, 
which has been used to evaluate the accuracy of molecular mechanics 
force fields and water models alike~\cite{Jorgensen2005PNAS}.   
%Figure \ref{solvation} shows that the hydration free energy 
%calculated by the OPC model
%is indeed more accurate than those calculated by the other
%commonly used models (TIP3P and TIP4PEw). 
To evaluate OPC's accuracy, we use a set of 20 molecules 
randomly selected to cover    
a wide range of experimental hydration energies from a 
large common  test set of small molecules~\cite{Mobley2009}, see \emph{Methods}.
Compared to experiment, OPC 
predicts hydration free energy more accurately, on average
(RMS error =  0.97 kcal/mol), 
as compared to 1.10 kcal/mol and 1.15 kcal/mol 
for TIP3P and TIP4PEw, respectively (Figure \ref{solvation}). 
The improvement is uniform across the range of solvation energies studied. 
The calculated average errors for OPC, TIP3P and TIP4PEw are 0.62, 0.78 and 0.87 kcal/mol, respectively, 
which shows that OPC is systematically more accurate than the other 
models tested.  OPC is more accurate despite the fact that force fields have been historically parametrized against TIP3P.
Somewhat paradoxically, TIP3P, which is certainly not the most accurate commonly used rigid model (see Figure~\ref{barchart}), 
has nevertheless been generally known thus far to give the highest accuracy in hydration free energy calculations \cite{Mobley2009}. 
The accuracy improvement by OPC is then noteworthy as it shows that 
an improvement in the ``right direction" can indeed lead to improvement 
in free energy estimates.     
To the best of our knowledge, OPC is the only classical rigid model 
that predicts the solvation free energies 
of small molecules with the ``chemical accuracy'' (RMS error $\leq$ 1 kcal/mol). % \cite{Nicholls2008}.  

\section{Concluding Remarks}

We have proposed a different approach to constructing classical water models.
This approach recognizes that commonly used distance and angle constraints 
on the configuration of a model's point charges are of little relevance to classical rigid water models; these artificial constraints complicate 
and impede the search for 
optimal charge distributions, key to reproducing unique features of liquid 
water. In our approach, such constraints are completely abandoned 
in favor of finding an optimal charge distribution (obeying only the fundamental C$_{2v}$ symmetry of water molecules) that best approximates properties of liquid water.

Next, we focus on the lowest multipole moments which directly control the electrostatics of the model. The hierarchical importance of these moments for 
water properties allowed us to reduce the search space to essentially 
just two key parameters: the dipole and the 
square quadrupole ($\mu$ and $Q_{T}$) moments; 
the less important moments 
%($Q_{0}$, $\Omega_{0}$ and $\Omega_{T}$) 
were 
fixed to the QM-derived values. 
The low dimensionality of the parameter space, combined with 
a set of derived equations that connects the geometrical and charge values to
its multipole moments,
 permitted a fine-grain exhaustive search virtually guaranteed 
to find an optimal solution within the accuracy class of the models considered here. The geometry of 
the resulting 4-point model (OPC) is different from commonly used ones, its 
location in the ($\mu$ and $Q_{T}$) is also distinctly different, 
which may  
explain why this optimum was not found previously via constrained
optimization in higher dimensional space of geometrical parameters.   
%By construction, the dipole, quadrupole and octupole moments of the 4-point rigid OPC model, determined by this approach, are more 
%consistent with quantum calculations, while other models 
%have much smaller dipole and quadrupole. 
The proposed  model 
is significantly more accurate than other commonly used rigid models in reproducing  bulk properties of liquid water. 
Although the optimization targeted a small 
subset of the properties at ambient conditions, 
the model reproduces a large number of bulk properties 
over a wide range of temperatures. 
%the temperature dependence of various properties of OPC is 
%significantly more accurate than the other commonly used models. 
The accuracy of predicted hydration free energies of small molecules 
has also improved, which is by no means an expected result: until 
now TIP3P model outperformed better quality rigid models in this respect. 
%it is well known
%that water models of lower over-all quality can still perform better 
%in these predictions. 
The consistently 
better accuracy offered by OPC demonstrates 
the benefits of fine-tuning electrostatic characteristics of classical 
water models.  
%for biomolecular simulations. 
%yet OPC  is significantly different in its parameters and geometry.             

We believe that the general approach presented here can be used  
to develop water models with different numbers
of point charges, including presumably even more 
accurate $n$-point ($n > 4$) models, 
and also flexible and polarizable models. We expect that 
finding an $n$-point charge optimum in the 2D parameter 
space ($\mu$, $Q_{T}$) is not going to be significantly more difficult than 
for the 4-point model presented here. 
The current 4-point OPC model is included in the solvent library of the 
Amber v14 molecular dynamics (MD) software package, 
and has been tested in GROMACS 4.6.5. 
The computational cost of running
molecular dynamics simulations with it is the same as that for the 
popular TIP4P model.

\section{Methods}
\subsection{Analytical solution for optimal point charges}   
%General procedure to compute optimal point charge approximations for a given set of multipole moments is outlined in \cite{Anandakrishnan2013}.
 Here we introduce the analytical equations that yield the positions and 
values of the three point charges that best reproduce the three lowest order multipole moments of the water molecule. 
% Assume that the elements of the dipole $\textbf{p}_{i}$, quadrupole $\textbf{Q}_{ij}$ and octupole $\textbf{O}_{ijk}$ tensors of the water molecule are %can be written as 
The lowest three nonzero multipole moments of the water molecule are the dipole that is represented by one independent component ($\mu$), 
the quadrupole defined by two independent components ($Q_{0}$, $Q_{T}$), and the octupole defined by two independent components ($\Omega_{0},\Omega_{T}$)~\cite{stone1997}.   
In the coordinate system shown in Figure~\ref{geoparamfig}, 
these moments are related to the Cartesian components of the traceless 
multipole moments of water molecule as 
$\mu = \mu_{z}$, 
$Q_{0} = Q_{zz}$, 
$Q_{T} =1/2(Q_{yy}-Q_{xx})$, 
${\Omega}_{0} = {O}_{zzz}$, and
$\Omega _{T}=1/2(O _{yyz}-O _{xxz})$~(see SI)~\cite{stone1997,Rick2004TIP5PEW,Niu2011Large}.

  The optimal point charges are calculated so that 
these moments are sequentially reproduced, starting with the lowest order moments \cite{Anandakrishnan2013}. 
The dipole and the quadrupole moments are reproduced exactly by requiring 
   \begin{eqnarray}
%     0 & = & q+q-2q   \\  
     \mu & = & 2q(z_{2}-z_{1}) \label{equdipole} \\  
      Q_{0} & = & -2q(\frac{y^2}{2}-z_{2}^2+z_{1}^2) \label{equQ0} \\ 
      Q_{T} &=& \frac{3qy^2}{2}  \label{equQt}
%      \Omega_{0}  &=& -2q(\dfrac{3}{2}y^2z_{2}-z_{2}^3+z_{1}^3)  \\  \label{equO0}
%      \Omega_{T} &=& \dfrac{5qy^2z_{2}}{2}   \label{equOt}
  \end{eqnarray}
   where $z_{2},z_{1},y$ and $q$ are the independent unknown parameters that characterize the three point charge model (see Figure \ref{geoparamfig}). 
   The above set of equations is solved to find three geometrical parameters of the water model ($z_{2},z_{1}$ and $y$) %:
 %  $z_{1,2} = (2Q_{T}+3Q_{0})/(6\mu) \mp \mu/4q$ and $y = \sqrt{2Q_{T}/(3q)}$.  

\begin{eqnarray}
z_{1,2}&= &(2Q_{T}+3Q_{0})/(6\mu) \mp \mu/4q \\
 y &=&\sqrt{2Q_{T}/(3q)}
\end{eqnarray}

This leaves only one unknown parameter, the charge value q, which we calculate by using two additional equations that 
    relate the charge distribution parameters to the octupole moment components so that 
    the octupole moment is optimally reproduced~\cite{Anandakrishnan2013}~(see SI). 

\subsection{Calculation of bulk properties}
The calculations of thermodynamic and dynamical bulk properties were done based on standard equations in the literature (see SI for details).  
Unless specified otherwise, we use the following Molecular Dynamics (MD) simulations protocol. 
Simulations in the NPT ensemble (1 bar, 298.16 K) were carried out using the Amber suite of programs. 
A cubic box with edge length of 30\AA~was filled with 804 water molecules. Periodic boundary condition was implemented in all directions. 
Long-range electrostatic interactions, calculated via the particle mesh Ewald (PME) summation, and the van der Waals interactions were cut off at distance 8\AA~. 
Dynamics were conducted with a 2~fs time step and all intra-molecular geometries were constrained with SHAKE. 
The NPT simulations were performed using Langevin thermostat with a coupling constant 2.0 $ps^{-1}$ and a Berendsen barostat with coupling constant of 1.0 $ps^{-1}$ for equilibration and 3.0 $ps^{-1}$ for production. 
The duration of production runs vary between 1~ns to 65~ns, depending on the properties (see SI). 

\subsection{Solvation free energy calculations}To avoid uncertainties due to conformational variability, the 20 test molecule 
were randomly selected from a subset of 248 highly 
rigid molecules~\cite{Mukhopadhyay2014}. 
Explicit solvent free energies calculations (via Thermodynamic Integration) 
were performed
in GROMACS 4.6.5 \cite{Pronk2013Gromacs} 
using the GAFF \cite{Wang2004GAFF} small molecule parameters, see SI 
for further details.

\subsection{Scoring function}
The predictive power of models against experimental data was validated using a scoring system developed by Vega et. al. \cite{Vega2011}.
% scores assign to the water  assigning a certain point according to the proximity to the experimental values (zero for poor agreement and ten for perfect agreement). We use the scoring system described by Vega et al (\cite{Vega2011}); 
For a calculated property $x$ and a corresponding experimental value of $x_{exp}$, the assigned score is obtained as~\cite{Vega2011} %,
%$M=max\lbrace[10 - |(x-x_{exp})\times100/(x_{exp}tol)|],0\rbrace$. 
%M=max\lbrace anint[10 - abs(\dfrac{(X-X_{exp})\times100}{X_{exp}tol})],0\rbrace
\begin{equation}
M=max\lbrace[10 - |(x-x_{exp})\times100/(x_{exp}tol)|],0\rbrace
\end{equation}
where the tolerance (tol) is assigned to 0.5\% for density,
position of the first peak in the RDF and heat of vaporization,
5\% for height of the first peak in the RDF,
and 2.5\% for the remaining properties. 
The quality score assigned to each test model is equal to the average of the scores in bulk properties considered.

%% supp
\section{Supplementary Materials}

\subsection{Analytical solution for optimal point charges}

%General procedure to compute optimal point charge approximations for a given set of multipole moments is outlined in \cite{Anandakrishnan2013}.
Here we present the analytical equations to find three point charges that optimally reproduce the dipole, the quadrupole and the octupole moments of the water molecule. 
In the coordinate system shown in Fig. 2 (main text), the elements of the traceless dipole $\textbf{p}_{i}$, quadrupole $\textbf{Q}_{ij}$ and octupole $\textbf{O}_{ijk}$ tensors \cite{Anandakrishnan2013} are %can be written as 
%    \begin{equation}
%      \textbf{q}=0
%      \label{monopoltensor}
%   \end{equation}
   \begin{equation}
      \textbf{p}_{i}=(0,0,\mu)
      \label{dipoletensor}
   \end{equation} 
   \begin{equation}
      \textbf{Q}_{ij}=
       \left( \begin{array}{ccc}
-Q_{T}-{Q_{0}}/{2} & 0 & 0 \\
0 & Q_{T}-{Q_{0}}/{2} & 0 \\
0 & 0 & Q_{0} \end{array} \right)
     \label{quadrupoletensor}
\end{equation}  
   \begin{equation}
      \textbf{O}_{ijk}=
       \left( \begin{array}{ccc}
-\Omega_{T}-{\Omega_{0}}/{2} & 0 & 0 \\
0 & \Omega_{T}-\Omega_{0}/{2} & 0 \\
0 & 0 & \Omega_{0} \end{array} \right)
    \label{octupoletensor}
\end{equation}
where $i, j = x, y$ and $k = z$, and 
$\mu,Q_{0},Q_{T}, \Omega_{0}$ and $\Omega_{T}$ are the dipole, 
the linear component of the quadrupole, the square component of the quadrupole,  
the linear component of the octupole, the square component of the octupole, respectively \cite{stone1997,Niu2011Large}. 
The other elements of the octupole tensor ($k = x, y$) can be found by symmetry.  
  % Assume that the positions and charge magnitudes of the three point charges are ($r_{1}=(0,-y,z_{2}),r_{2}=(0,y,z_{2}),r_{3}=(0,0,z_{1})$) and ($q_{1}=q,q_{2}=q, q_{3}=-2q$) (Fig. \ref{geoparam}), 
The optimal charge values and positions are calculated so that 
these three moments are sequentially reproduced, starting with the lowest order moments \cite{Anandakrishnan2013}. 
The first two lowest order moments of the water molecule, the dipole and the quadrupole, are fully reproduced by requiring 
   \begin{eqnarray}
%     0 & = & q+q-2q   \\  
     \mu & = & 2q(z_{2}-z_{1}) \label{equdipole} \\  
      Q_{0} & = & -2q(\frac{y^2}{2}-z_{2}^2+z_{1}^2) \label{equQ0} \\ 
      Q_{T} &=& \frac{3qy^2}{2}  \label{equQt}
%      \Omega_{0}  &=& -2q(\dfrac{3}{2}y^2z_{2}-z_{2}^3+z_{1}^3)  \\  \label{equO0}
%      \Omega_{T} &=& \dfrac{5qy^2z_{2}}{2}   \label{equOt}
  \end{eqnarray}
   where $z_{2},z_{1},y$ and $q$ are independent unknown parameters that characterize the three point charge model (see Fig. 2). 
   The above three equations are solved to find three geometrical parameters ($z_{2},z_{1}$ and y), as follows  
     \begin{eqnarray}
     &&  z_{1,2} = \frac{2Q_{T}+3Q_{0}}{6\mu} \mp \frac{\mu}{4q}  \label{OPCsolutions1} \\ 
      && y = \sqrt{\frac{2Q_{T}}{3q}}       \label{OPCsolutions2}
      \end{eqnarray}
      
   For a given value of q, the values of $z_{2},z_{1}$ and $y$ found as above 
   exactly reproduce the dipole ($\mu$) and the quadrupole ($Q_{0}$ and $Q_{T}$) moments of interest. 
   The only remaining unknown parameter, q, is found to optimally reproduce  
   the next order moment, the octupole, which is described by two independent parameters ($\Omega_{0}$ and $\Omega_{T}$).
   The components of the octupole moment are related to the charge distribution parameters through
   
  \begin{eqnarray}
     \Omega_{0}  & = & -2q(\frac{3}{2}y^2z_{2}-z_{2}^3+z_{1}^3) \label{equO0} \\ 
     \Omega_{T} & = & \frac{5qy^2z_{2}}{2} \label{equOt}
  \end{eqnarray}

   The octupole tensor (Eq.~\ref{octupoletensor}) can be optimally approximated if 
   the largest absolute principal value of the octupole tensor (i.e. $(\Omega_{T}-\Omega_{0}/2)$ for the water molecule) is reproduced \cite{Anandakrishnan2013}. 
   Therefore, we set ($\Omega_{T}-\Omega_{0}/2$) from Eqs. \ref{equO0} and \ref{equOt} and solve for q as

\begin{eqnarray}
  && q = -3\frac{\sqrt{\mu^{4}(256 Q_{T}^2 + \xi)}+16 Q_{T} \mu^2}{2\xi} \label{OPCsolutions3} \\     
  where \nonumber\\
  && \xi = 52 Q_{T}^2+60 Q_{T} Q_{0}-9 (3 Q_{0}^2+8 (\Omega_{T}-\Omega_{0}/2) \mu) \nonumber 
  \end{eqnarray}

The above solution is valid only when $\xi < 0$. For $\xi \geq 0$, the point charge positions converge to a singular point and the charge values go to infinity. The corresponding region in $\mu-Q_{T}$ map (Fig. 3) leading to this condition is displayed in deepest red (zero score).

%\section{Supporting Information} 
\subsection{Solvation free energy calculations} 
Standard thermodynamics integration (TI) protocol was adopted from Ref.~\cite{Mobley2009}. 
The Merck-Frosst implementation of AM1-BCC \cite{Jakalian2000,Jakalian2002} was used to assign the partial charges.
The topology and coordinates for the
molecules were obtained from Ref. \cite{Mobley2009}. Molecules were solvated
in triclinic box with at least 12 \AA\ from the solute to the nearest
box edge. After minimization and equilibration,
we performed standard free energy perturbation calculations
using 20 $\lambda$ values. Real space electrostatic cutoff was 10 \AA.
All bonds were restrained using the LINCS algorithm. Production NPT simulations were performed for 5ns. Identical
simulations were performed for TIP3P, TIP4PEw, and OPC.

\subsection{Calculating the bulk properties}

The calculation of bulk properties were done based on standard equations in the literature \cite{TIP4PEW,Voth07,TIP4P2005,TIP4Peps}. 
Unless stated otherwise, values of OPC at ambient temperature (Table 3) are given 
as averages over six independent simulations of 65 ns each, 
except for those quantities that are derived from temperature dependent results. 
The temperature dependent results are calculated from one simulation of 65 ns for each temperature point, i.e. 12.5K intervals in a temperature range [248K, 373K]. 
Details of the calculations of studied quantities are described below.

%\setcounter{figure}{0}
%\makeatletter 
%\renewcommand{\thefigure}{S\@arabic\c@figure}
%\makeatother
%\setcounter{equation}{0}
%\section{Equations to calculate the bulk properties}
%The calculation of thermodynamic and dynamical bulk properties were done based on the standard equations. % explained in the Methods section \cite{Horn2004}. 
\subsubsection{Static dielectric constant}The static dielectric constant $\epsilon_{0}$ is determined through \cite{TIP4PEW,TIP4P2005,TIP4Peps} 
 
\begin{equation}
\epsilon_{0} = 1+ \frac{4\pi}{3k_{B}TV}(<\textbf{M}^{2}>-<\textbf{M}>^{2})
\label{epsequ}
\end{equation}
where $\textbf{M}=\Sigma_{i}q_{i}\textbf{r}_{i}$, $\textbf{r}_{i}$ is the position of atom i, $k_{B}$ is the Boltzmann constant, $T$ is the absolute temperature and $V$ is the simulation box average volume.  
%Simulation length over 65 ns was required for the quantity to converge. %Results from the first 20~ns trajectories were excluded from the averaging~\cite{Karimian2011}. 
%The reported value at ambient condition is an average over six independent simulations.

\subsubsection{Self diffusion coefficient}The self-diffusion coefficient $D$ is obtained using the Einstein relation \cite{TIP4PEW,TIP4Peps,Voth07} 
 
\begin{equation}
D=\lim_{t \,\rightarrow\, \infty} \frac{1}{6t}<|r(t)-r(0)|^{2}>
\label{diffequ}
\end{equation}
%\begin{figure}[h]
%\centerline{\includegraphics[width=1.0\linewidth]{diff_temp.eps}}
%\caption{Figure caption}\label{placeholder}
%\end{figure}

The simulation protocol to compute the self-diffusion coefficient is similar to the protocol described in Ref.~\cite{TIP4PEW}; the well equilibrated NPT simulations were followed up with 80 successive intervals of NVE (20 ps) and NPT (5 ps) ensembles. The self diffusion was obtained by averaging $D$ values over all the NVE runs. %The reported value at ambient condition is an average over six independent simulations. 

\subsubsection{Heat of vaporization} The heat of vaporization $\Delta H_{vap}$ is obtained following the method described in Ref. \cite{TIP4PEW}, as

\begin{equation}
\Delta  H_{vap} \approx - U_{liq}/N+RT-pV-E_{pol}+C
\label{hvapequ}
\end{equation}
where $U_{liq}$ is the potential energy of the liquid with $N$ molecules at a given external pressure $p$ and a temperature $T$, and $V$ is the average volume of the simulation box. $R$ is the ideal gas constant. $E_{pol}$ accounts for the energetic cost of the effective polarization energy, and can be approximated as

\begin{equation}
 E_{pol}=\frac{(\mu-\mu_{gas})^{2}}{2\alpha_{gas}} 
 \label{epolequ}
\end{equation}
 where $\mu$ is the dipole moment of the corresponding rigid model and $\mu_{gas}$ and $\alpha_{gas}$ are the dipole moment and the mean polarizability of a water molecule in the gas phase \cite{guidline2001}, respectively. 
 The OPC's dipole is close to experiment and larger than that of common rigid models which 
 yields a relatively larger value of $E_{pol}$ for OPC compared to common rigid models.  
 The correction term C, which accounts for vibrational, nonideal gas, and pressure effect, for various temperatures is taken from Ref.~\cite{TIP4PEW}.

\subsubsection{Isobaric heat capacity}
The isobaric heat capacity $c_{p}$ is determined through numeric differentiation of simulated enthalpies $H(T)$ over the range of temperatures $T$ of interest~\cite{TIP4PEW,Voth07}

\begin{equation}
C_{p}\approx\frac{<H(T_{2})>-<H(T_{1})>}{T_{2}-T_{1}}+ \Delta C_{QM}
\label{cpequ}
\end{equation}
where $\Delta C_{QM}$ ($\approx -2.2408$ at $T=298.0 K$) is a quantum correction term accounting for the quantized character of the neglected intramolecular vibrations. The values of $\Delta C_{QM}$ for different temperatures are taken from Ref.~\cite{TIP4PEW}. The numeric differentiation is calculated from simulations in the temperature range [248K, 373K] in 12.5K increments.

\subsubsection{Thermal expansion coefficient}
The thermal expansion coefficient $\alpha_{p}$ can be approximated through numeric differentiation of simulated bulk-densities $\rho(T)$ over a range of temperatures $T$ of interest~\cite{TIP4PEW,Voth07,TIP4Peps}

\begin{equation}
\alpha_{p} \approx -(\frac{\ln<\rho(T_{2})>-\ln<\rho(T_{1})>}{T_{2}-T_{1}})_{P}
\label{alphapequ}
\end{equation}

The reported value at ambient conditions is calculated from a numeric differentiation of bulk-densities at $T_{1}$=296K and $T_{2}$=300K, averaged over 4 independent simulations.

\subsubsection{Isothermal compressibility}
The isothermal compressibility $\kappa_{T}$ is calculated from volume fluctuations in NPT simulation 
using a Langevin thermostat with coupling constant 2.0 $ps^{-1}$ and a Monte Carlo barostat with coupling constant of 3.0 $ps^{-1}$, 
via the following formula~\cite{TIP4PEW,TIP4P2005,TIP4Peps}

\begin{equation}
\kappa_{T} = \frac{<V^{2}>-<V>^{2}}{k_{B}T<V>}
\label{kappatequ}
\end{equation}
% \kappa_{T} \approx -(\frac{\ln<\rho(P_{2})>-\ln<\rho(P_{1})>}{P_{2}-P_{1}})_{T}

%Simulation length over 65 ns was required for the quantity to converge. 
Simulations of 65ns and 15ns time length were performed to obtain the temperature dependent results for ($T\leq298K$) and ($T>298K$), respectively. 
%The temperature dependence results are obtained through an order three polynomial fit of averaged results from two independent simulations at studied temperature. 

\subsubsection{Propensity for Charge Hydration Asymmetry}
Propensity of a water model to cause Charge Hydration Asymmetry (CHA) for a similar size cation/anion 
pair ($B^{+}/A^{-}$) such as $K^{+}/F^{-}$ is defined in Ref.~\cite{Mukhopadhyay2012Charge} as 

\begin{equation}
\eta^{*}(B^{+}/A^{-})=\frac{\Delta G(B^{+}) - \Delta G(A^{-})}{1/2|\Delta G(B^{+}) + \Delta G(A^{-})|} \approx 2\frac{\frac{\widetilde{Q}_{zz}}{\mu}}{R_{iw}}
\label{CHA}
\end{equation}
where the term on the right is an approximation of propensity for CHA for point charge water models ~\cite{Mukhopadhyay2012Charge}, 
$R_{iw}$ is the ion-water distance, $\Delta G$ is the free energy of hydration, and $\mu$ and $\widetilde{Q}_{zz}$ are the dipole and the nontraceless quadrupole moment of the model, respectively~\cite{stone1997}.

%\pagebreak

\subsection{Additional bulk properties, comparison with most recent models}
\subsubsection{O-O radial distribution function}
\begin{figure}[!h]
%\centerline{\includegraphics[width=1\linewidth]{RDF.eps}}
\centering
\includegraphics[width=0.41\linewidth]{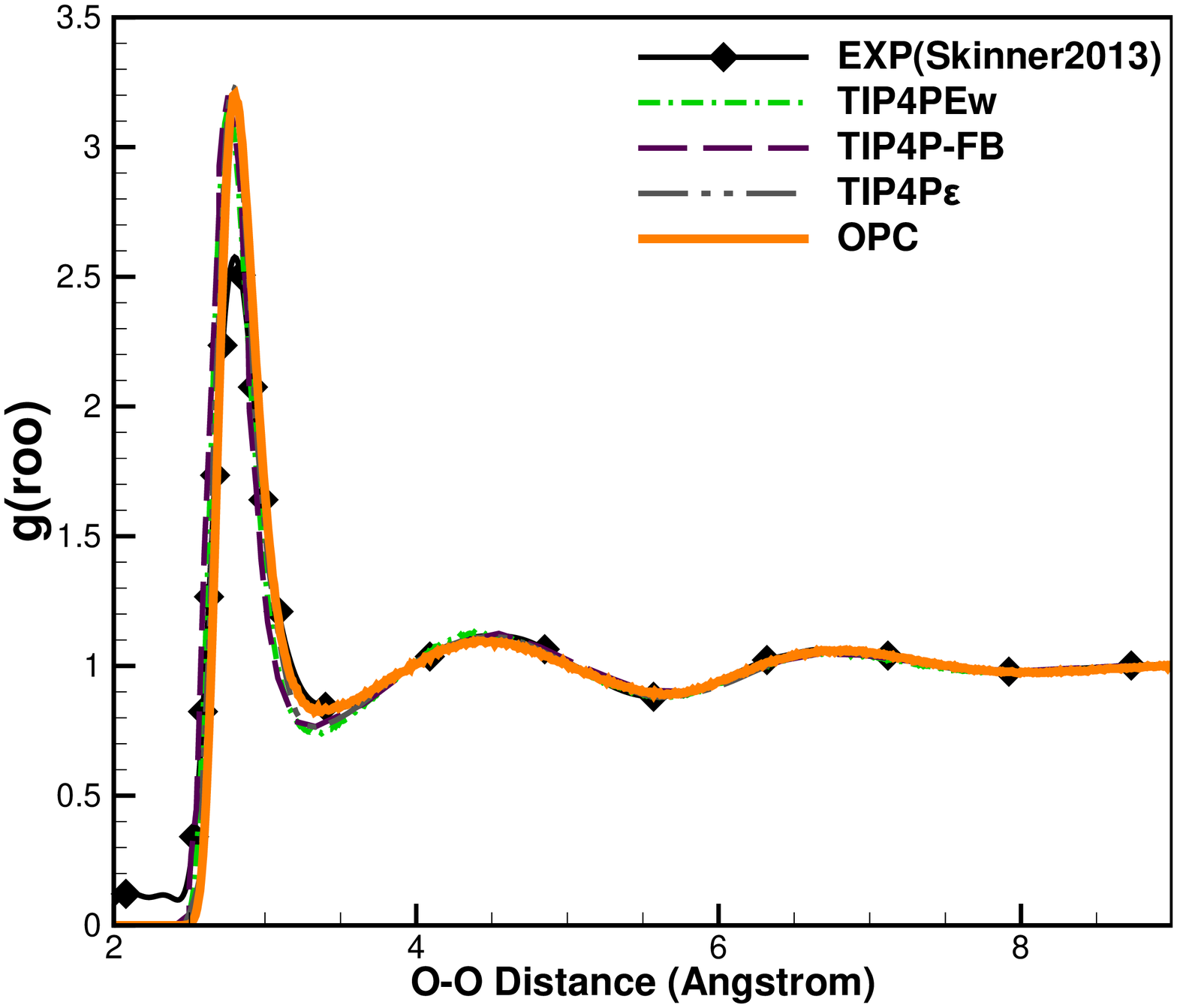}
\includegraphics[width=0.41\linewidth]{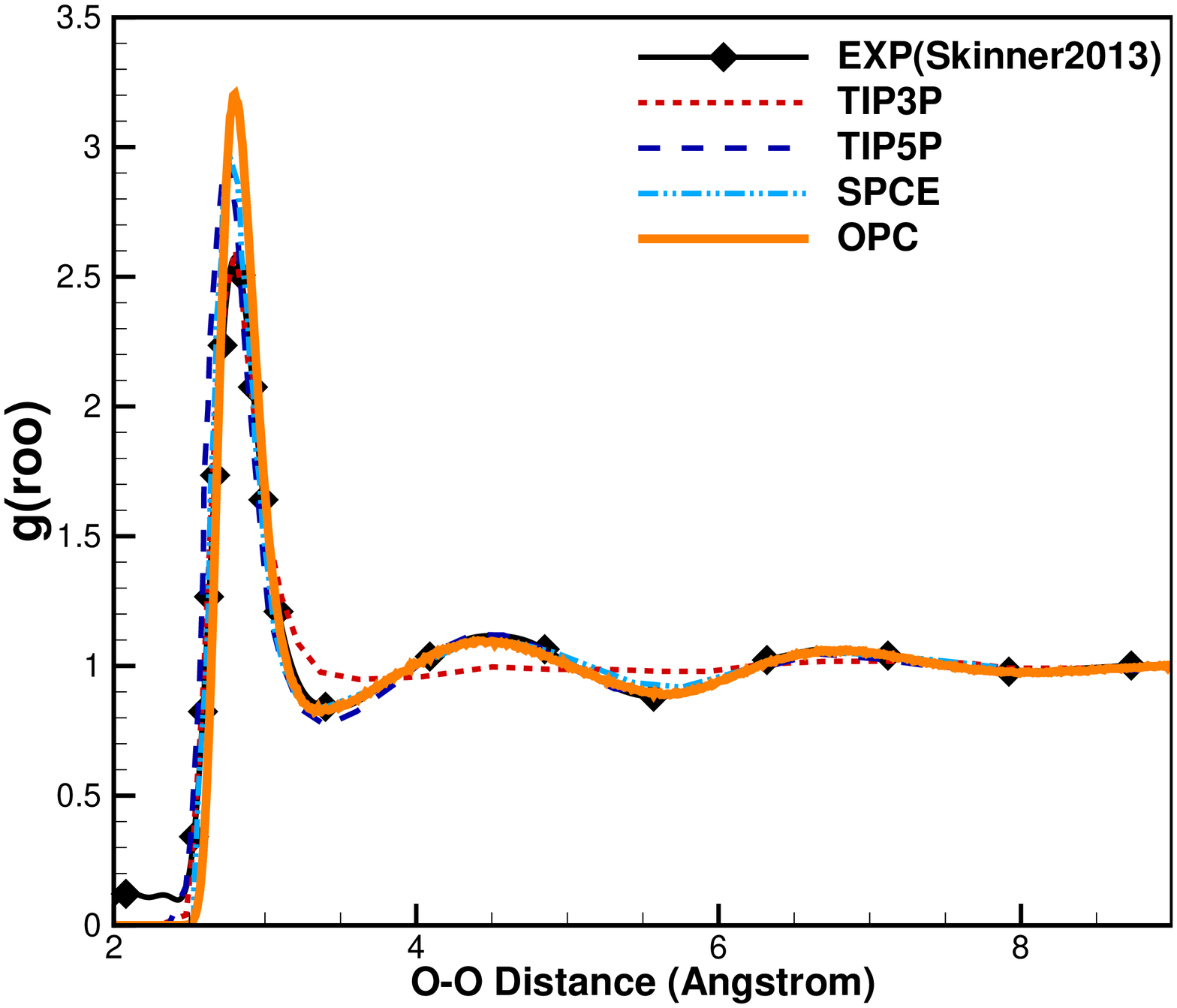}
%\centerline{\includegraphics[width=1\linewidth]{RDF5p.eps}}
\caption{O-O radial distribution function of liquid water at 
298.16 K, 1 bar. The OPC model is compared to the commonly used rigid  models as well as some recent rigid models (TIP4P-FB and TIP4P$\epsilon$). 
The experimental data is taken from \cite{Skinner2013}. TIP4PEw result is from \cite{TIP4PEW}, TIP4P-FB from \cite{Wang2014pande}, 
TIP4P$\epsilon$ from \cite{TIP4Peps}, SCPE from \cite{SPCE}, TIP3P from \cite{Jorgensen1998tip3pspctip4p} and TIP5P from \cite{TIP5P}.}\label{rdfequ}
\end{figure}

Each potential OPC model is parametrized to exactly reproduce the position of the first peak. 
% [[[while common models are calibrated with the help of old and possibly less accurate X-ray data that normally underestimate the position of the first peak]]] (\ref{table:bulkproperties}). 
The positions and the heights of the remaining peaks are very accurately reproduced with these parameters. 
The height of the first peak is however slightly high,
which leads to an average O-O coordination number ($n_{oo}$) larger than 
experiment. 
This may be because of the $r^{-12}$ repulsion 
in the LJ potential 
that is known to create an over structured liquid \cite{Kiss2013,Guillot2002}. 
It is argued that using a softer potential 
(e.g. a simple exponential in the form of $A e^{âˆ’B_r}$) can correct the height of the first peak \cite{Kiss2013}. 
% [[[This is because the decay of the electron density is exponential at high distances and accordingly an exponential function it essential to provide a more realistic description of the repulsion between adjacent molecules as a result of the overlapping electron clouds]]]. 
 We employ a 12-6 potential to achieve compatibility with standard biomolecular force fields.
While TIP3P is the only model that accurately reproduces the height of the first peak, it lacks structure beyond the first coordination shell (Fig. \ref{rdfequ}). 
%as indicated by the ?lack? of a second RDF peak.
% The TIP4P-Ew model presents a relatively good structure, 
% but looses accuracy in many other properties, 
% as seen in tabel \ref{table:bulkproperties}. 

\pagebreak

\subsubsection{Isobaric heat capacity, isothermal compressibility, recent models}  

\begin{figure}[!h]
%\FloatBarrier
%\vspace*{-.3in}
\centering
\includegraphics[width=.41\linewidth]{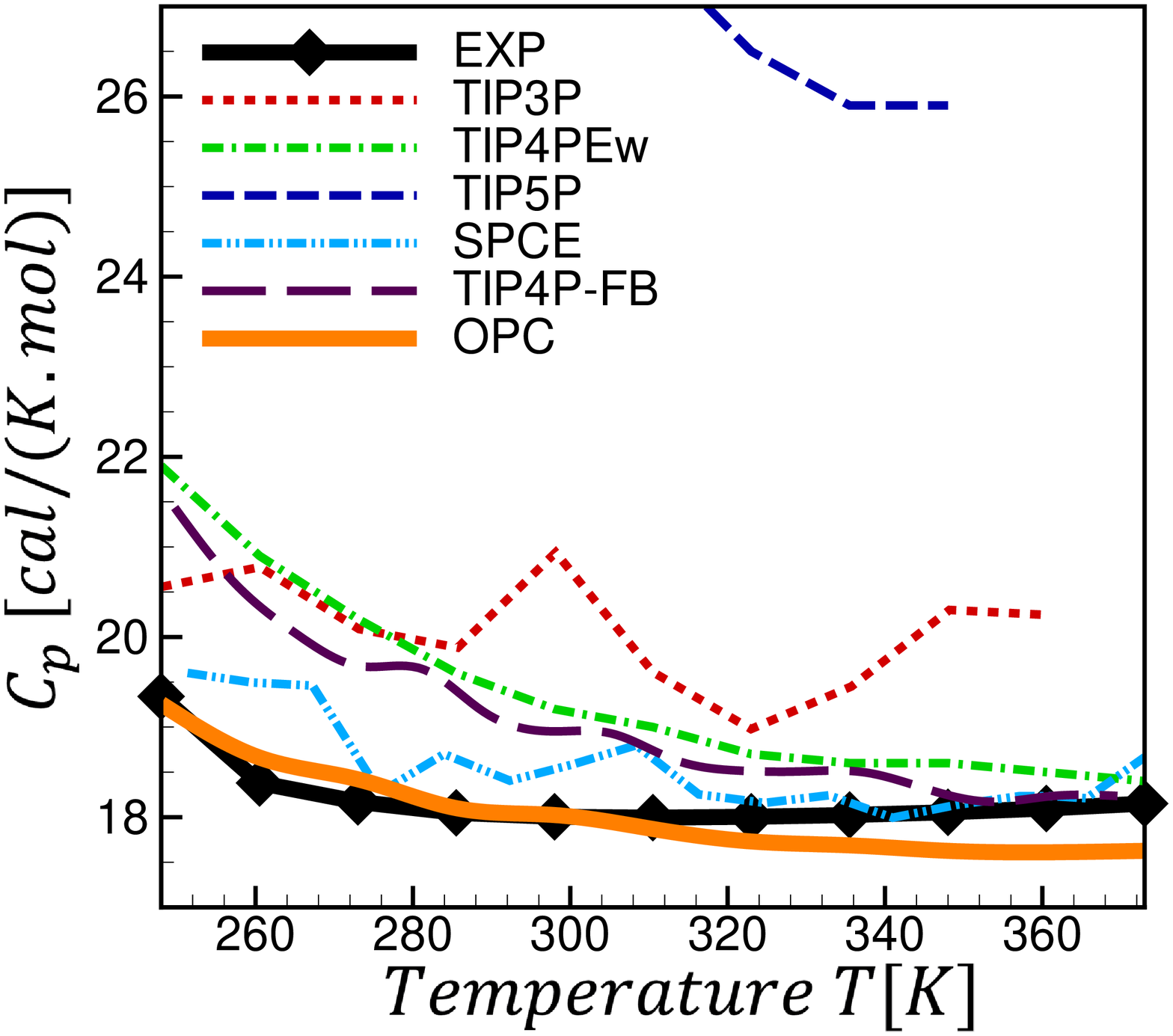}
\includegraphics[width=.41\linewidth]{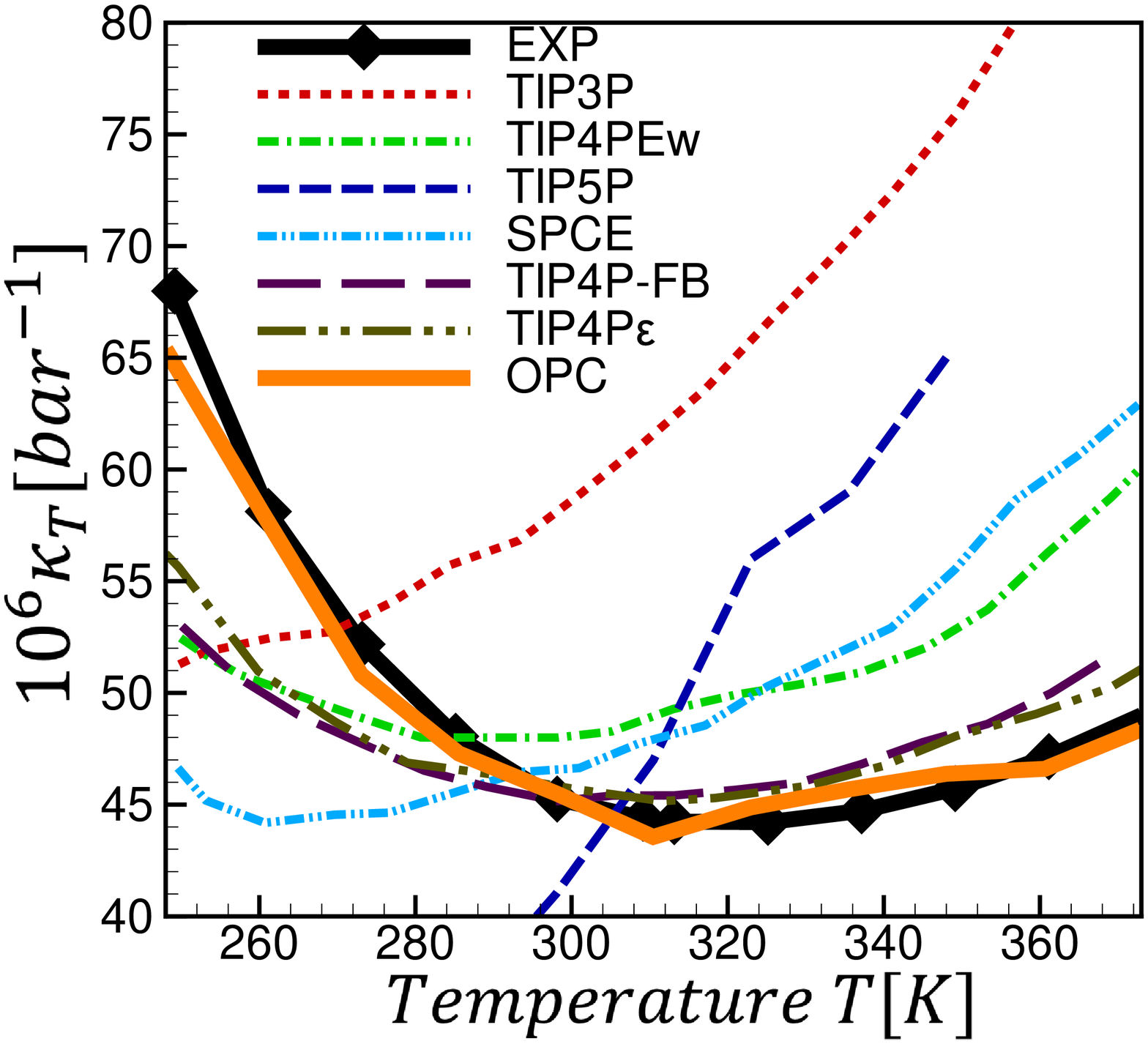}
%(d)\\
\caption{ Variation of isobaric heat capacity and isothermal compressibility of liquid phase water with temperature. 
OPC model (this work) is compared to several common rigid models, some recent rigid models (TIP4P-FB and TIP4P$\epsilon$) and experiment. 
TIP4PEw results are from \cite{TIP4PEW}, TIP5P from \cite{TIP5P}, TIP3P from \cite{TIP3P,Wang2014pande}, 
SPCE and TIP4P-FB from \cite{Wang2014pande}, and TIP4P$\epsilon$ from \cite{TIP4Peps}. }\label{tempdepcp} 
\end{figure}

%\pagebreak
\begin{figure}[t]
%\vspace*{1in}
\centerline{\includegraphics[width=.9\linewidth]{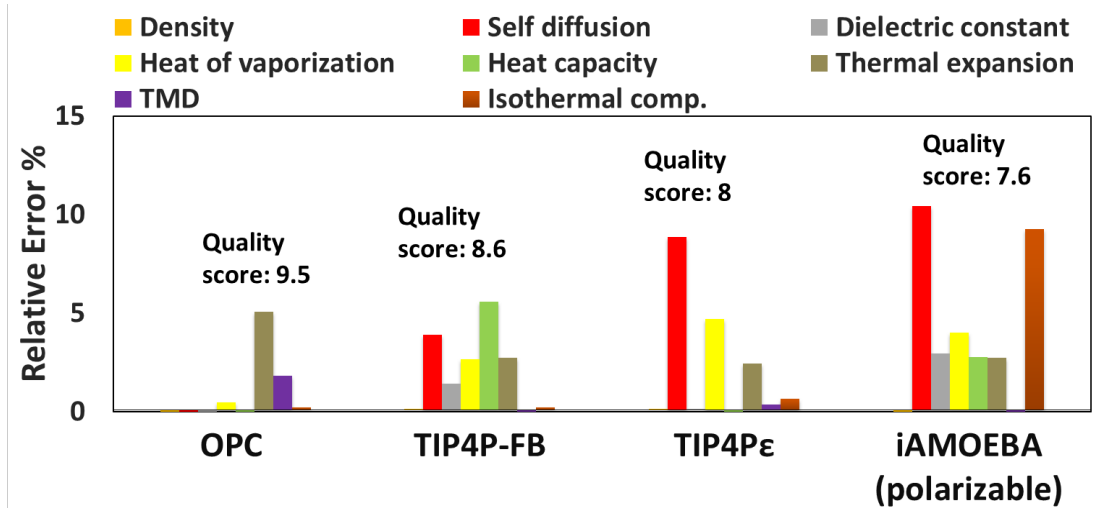}}
\caption{Comparing the accuracy of OPC to some recent rigid water models (TIP4P-FB \cite{Wang2014pande} and TIP4P$\epsilon$ \cite{TIP4Peps}), 
including a polarizable one (iAMOEBA \cite{Wang2013iAMOEBA}). 
The quality scores (see \emph{Methods}) represent the overall performance of each model in reproducing eight key properties, i.e. 
density $\rho$, self diffusion coefficient $D$, static dielectric constant $\epsilon_{0}$, 
heat of vaporization $\Delta H_{vap}$, isobaric heat capacity $C_{p}$, isothermal compressibility $\kappa_{T}$ and
thermal expansion coefficient $\alpha_{p}$, at ambient conditions, as well as the temperature of maximum density (TMD). 
The heat capacity value for TIP4P$\epsilon$ is not reported in the original reference \cite{TIP4Peps}, 
and therefore was excluded from the quality score calculated for this model.}\label{recentmodels}
\end{figure}

\clearpage

%% supp

\begin{acknowledgement}
%%Support from  is acknowledged. 
This work was supported by NIH GM076121, and in part by NSF grant CNS-0960081 and the HokieSpeed supercomputer at Virginia Tech. 
We thank Lawrie B. Skinner and Chris J. Benmore for providing experimental oxygen-oxygen pair-distribution 
function of water. 

\end{acknowledgement}

%\begin{suppinfo}
%
%Analytical solution for optimal point charges, 
%detailed procedure for calculating bulk properties and solvation free energies, 
%additional bulk properties and comparison with most recent water models. \end{suppinfo}
%\bibliographystyle{achemso}
%\bibliography{watermultipole.jacs}

\providecommand{\latin}[1]{#1}
\providecommand*\mcitethebibliography{\thebibliography}
\csname @ifundefined\endcsname{endmcitethebibliography}
  {\let\endmcitethebibliography\endthebibliography}{}

\end{document}